\documentclass[twocolumn]{aastex631}

\usepackage{longfigure}
\usepackage{amsmath}
\usepackage[version=4]{mhchem} 

\DeclareRobustCommand{\ion}[2]{%
\relax\ifmmode
\ifx\testbx\f@series
{\mathbf{#1\,\mathsc{#2}}}\else
{\mathrm{#1\,\mathsc{#2}}}\fi
\else\textup{#1\,{\mdseries\textsc{#2}}}%
\fi}

\shorttitle{Accurate Strong Lensing Model of SDSS J1029+2623}
\shortauthors{Acebron et al.}
\graphicspath{{./}{figures/}}
\begin{document}

\title{VLT/MUSE observations of SDSS J1029+2623: towards a high-precision strong lensing model\footnote{This work is based in large part on data collected at ESO VLT (prog. ID 0102.A-0642(A)) and NASA \textit{HST}.}}

\correspondingauthor{Ana Acebron}
\email{ana.acebron@unimi.it}

%%%%%%%%%%%%%%%%%%%%%%%%%%%%%%%%%%%%%%%%%%%%%%%
\author[0000-0003-3108-9039]{Ana Acebron}
\affiliation{Dipartimento di Fisica, Universit\`a degli Studi di Milano, Via Celoria 16, I-20133 Milano, Italy}

\author[0000-0002-5926-7143]{Claudio Grillo}
\affiliation{Dipartimento di Fisica, Universit\`a degli Studi di Milano, Via Celoria 16, I-20133 Milano, Italy}
%\affiliation{Dark Cosmology Centre, Niels Bohr Institute, University of Copenhagen, Jagtvej 128, DK-2200 Copenhagen, Denmark}

\author[0000-0003-1383-9414]{Pietro Bergamini}
\affiliation{INAF – OAS, Osservatorio di Astrofisica e Scienza dello Spazio di Bologna, via Gobetti 93/3, I-40129 Bologna, Italy}

\author[0000-0001-9261-7849]{Amata Mercurio}
\affiliation{INAF – Osservatorio Astronomico di Capodimonte, Via Moiariello 16, I-80131 Napoli, Italy}

\author[0000-0002-6813-0632]{Piero Rosati}
\affiliation{Dipartimento di Fisica e Scienze della Terra, Universit\`a degli Studi di Ferrara, via Saragat 1, I-44122 Ferrara, Italy}
\affiliation{INAF – OAS, Osservatorio di Astrofisica e Scienza dello Spazio di Bologna, via Gobetti 93/3, I-40129 Bologna, Italy}

\author[0000-0001-6052-3274]{Gabriel Bartosch Caminha}
\affiliation{Max-Planck-Institut f\"ur Astrophysik, Karl-Schwarzschild-Str. 1, D-85748 Garching, Germany}

\author[0000-0003-3096-9966]{Paolo Tozzi}
\affiliation{INAF – Osservatorio Astrofisico di Arcetri, Largo E. Fermi 5, I-50125, Firenze, Italy}

\author[0000-0003-2680-005X]{Gabriel B. Brammer}
\affiliation{Cosmic Dawn Center (DAWN), Jagtvej 128, DK2200 Copenhagen N, Denmark}
\affiliation{Niels Bohr Institute, University of Copenhagen, Jagtvej 128, København N, DK-2200, Denmark}

\author[0000-0003-1225-7084]{Massimo Meneghetti}
\affiliation{INAF – OAS, Osservatorio di Astrofisica e Scienza dello Spazio di Bologna, via Gobetti 93/3, I-40129 Bologna, Italy}
\affiliation{National Institute for Nuclear Physics, viale Berti Pichat 6/2, I-40127 Bologna, Italy}
\affiliation{Division of Physics, Mathematics, \& Astronomy, California Institute of Technology, Pasadena, CA 91125, USA}

\author{Andrea Morelli}
\affiliation{Dipartimento di Fisica e Scienze della Terra, Universit\`a degli Studi di Ferrara, via Saragat 1, I-44122 Ferrara, Italy}

\author[0000-0001-6342-9662]{Mario Nonino}
\affiliation{INAF – Osservatorio Astronomico di Trieste, via G. B. Tiepolo 11, I-34131 Trieste, Italy}

\author[0000-0002-5057-135X]{Eros Vanzella}
\affiliation{INAF – OAS, Osservatorio di Astrofisica e Scienza dello Spazio di Bologna, via Gobetti 93/3, I-40129 Bologna, Italy}

%%%%%%%%%%%%%%%%%%%%%%%%%%%%%%%%%%%%%%%%%%%%%%%

%%\collaboration{6}{(AAS Journals Data Editors)}




\begin{abstract}
We present a strong lensing analysis of the galaxy cluster SDSS J1029+2623 at $z=0.588$, one of the few currently known lens clusters with multiple images of a background ($z=2.1992$) quasar with a measured time delay. We use archival Hubble Space Telescope multi-band imaging and new Multi Unit Spectroscopic Explorer follow-up spectroscopy to build an accurate lens mass model, a crucial step towards future cosmological applications. 
The spectroscopic data enable the secure identification of 57 cluster members and of two nearby perturbers along the line-of-sight.
We estimate the inner kinematics of a sub-set of 20 cluster galaxies to calibrate the scaling relations parametrizing the sub-halo mass component. 
We also reliably determine the redshift of 4 multiply imaged sources, provide a tentative measurement for one system, and report the discovery of a new four-image system. The final catalog comprises 26 multiple images from 7 background sources, spanning a wide redshift range, from 1.02 to 5.06. 
We present two parametric lens models, with slightly different cluster mass parametrizations. The observed positions of the multiple images are accurately reproduced within approximately $0\arcsec.2$, the three image positions of the quasar within only  $\sim0\arcsec.1$. 
We estimate a cluster projected total mass of $M(<300~ {\rm kpc}) \sim 2.1 \times 10^{14}~ M_{\odot}$, with a statistical uncertainty of a few percent. Both models, that include a small galaxy close to one of the quasar images, predict magnitude differences and time delays between the quasar images that are consistent with the observations.
\end{abstract}




\section{Introduction} \label{sec:intro}
In the last years, thanks to the increased precision on the measurements of the value of the Hubble constant ($H_0$), some tension has emerged between the estimates from local and early-Universe probes \citep[][]{Riess2019, Riess2021, Aghanim2020}. This discrepancy \citep[see][for a review]{Verde2019} may point to the presence of unknown systematic effects \citep{Freedman2020} or towards a deviation from the standard flat $\mathrm{\Lambda CDM}$ model and therefore interesting new physics. However, an explanation for this discrepancy has yet to be found, making the exploration of additional independent, complementary and high-precision techniques fundamental.

Since Refsdal theoretically predicted that a strongly lensed supernova (SN) with measured time delays between its multiple images could provide an independent way to measure the value of the Hubble constant \citep{Refsdal1964}, this technique has proved to provide competitive estimates of the value of $H_0$.
However, given the rarity of lensed SNe \citep{Kelly2015, Goobar2017, Rodney2021}, the strong lensing time delay method has been mostly exploited with quasars strongly lensed by galaxies. In particular, the $H_0$ Lenses in COSMOGRAIL’s Wellspring (H0LiCOW) program \citep{Suyu2017}, together with the COSmological MOnitoring of GRAvItational Lenses (COSMOGRAIL) program \citep[e.g.,][]{Tewes2013, Courbin2018, Bonvin2018}, have estimated the value of $H_0$ with a $2.4\%$ precision from the joint analysis of six gravitationally lensed quasars with measured time delays \citep{Birrer2019, Sluse2019, Wong2020}.

Using time-varying sources strongly lensed by galaxy clusters \citep{Inada2003, Oguri2013, Sharon2017, Grillo2020} is a complementary technique that remains largely unexploited. On the one hand, galaxy-scale systems are more common, require a simpler total mass modeling, and may be less affected by line-of sight mass structures. On the other hand, cluster-scale systems are less prone to the so-called mass-sheet and mass-slope degeneracies (see e.g. \citealt{Grillo2020}), and the longer time delays allow for measurements with smaller relative uncertainties, that can reach a few percent precision.

On galaxy cluster scales, SN “Refsdal” was discovered by \citet{Kelly2015} to be strongly lensed by the Hubble Frontier Fields \citep{Lotz2017} galaxy cluster MACS J1149.5+2223 \citep{Treu2016,Grillo2016}. 
By exploiting Multi Unit Spectroscopic Explorer \citep[MUSE,][]{Bacon2010, Bacon2014} spectroscopic identifications of a large number of multiple images in MACS J1149.5+2223, \citet{Grillo2018} inferred the value of the Hubble constant from the first multiply-imaged and spatially-resolved SN Refsdal, using a full strong lensing analysis including the SN measured time delays \citep{Kelly2016, Rodney2016}. These first results suggest that time delays in lens galaxy clusters will become an important and complementary tool to measure the expansion rate and the geometry of the Universe.

As shown in \citet{Grillo2018,Grillo2020}, to take full advantage of such lens clusters (that yield measured time delays with a few percent precision) as cosmological probes, it is necessary to construct an accurate cluster total mass model.
In that sense, MUSE spectroscopy together with high-resolution Hubble Space Telescope (HST) imaging, has allowed in the past few years the development of high precision strong lensing (SL) mass models through the identification of a large number of multiple images \citep[e.g.,][]{Richard2015, Kawamata2016, Caminha2017, Mahler2018, Caminha2019, Jauzac2019, Rescigno2020, Lagattuta2019, Ghosh2021, Jauzac2021} and the use of stellar kinematics of cluster galaxies \citep[][]{Bergamini2019, Bergamini2021, Pignataro2021, Granata2021}. 

In this work, we present a new SL model of the galaxy cluster SDSS J1029+2623, at a redshift of $z=0.588$, based on recent MUSE observations.
SDSS J1029+2623, hereafter SDSS1029, is one of the few presently known lens clusters producing multiple images (3, labeled as A, B, and C) of a background ($z = 2.1992$) quasar \citep[][]{Inada2006, Oguri2008, Oguri2013} with a large maximum image separation of $ \sim 22\arcsec.5$ \citep{Inada2006}.
SDSS1029 was discovered in the Sloan Digital Sky Survey Quasar Lens Search \citep[SQLS,][]{Oguri2006, Inada2012}, which is a large survey that followed-up gravitationally lensed quasars that were spectroscopically confirmed in the Sloan Digital Sky Survey \citep[SDSS,][]{York2000}.
The first strong lensing analysis of the cluster, using deep multi-band HST observations, was presented in \citet{Oguri2013}, where they also identified several other plausible multiple image systems, all lacking a spectroscopic confirmation.

Moreover, after a 5.4 year long optical monitoring campaign, a time delay of $744~\pm~10$ days was measured between the images A and B of the lensed quasar \citep{Fohlmeister2013}. 
The claimed $\sim1\%$ uncertainty in this time delay offers an excellent opportunity for testing the cosmological applications of lens galaxy clusters with multiply imaged time-varying sources.

This paper is organized as follows. In Section \ref{sec:data}, we describe the HST imaging and MUSE spectroscopic observations. Section \ref{sec:SLM} presents the selection of the multiple images and cluster members, as well as the adopted methodology for the strong lensing modeling of SDSS1029. Our findings are presented and discussed in Section \ref{sec:results}. Finally, our conclusions are summarized in Section \ref{sec:conclu}. 

Throughout the paper, we adopt the standard ${\Lambda\mathrm{CDM}}$ flat cosmological model with $H_0 = 70$ $   \mathrm{km~s^{-1}~Mpc^{-1}}$, $\mathrm{\Omega_{m}}=0.3$ and $\mathrm{\Omega_{\Lambda}}=0.7$. In this cosmology, $1\arcsec$ corresponds to a physical scale of 6.62 kpc at the cluster redshift ($z=0.588$).
Magnitudes are quoted in the AB system. Statistical uncertainties are given as the 68\% confidence interval unless otherwise noted.

\begin{figure*}
\centering
\includegraphics[width=0.82\linewidth]{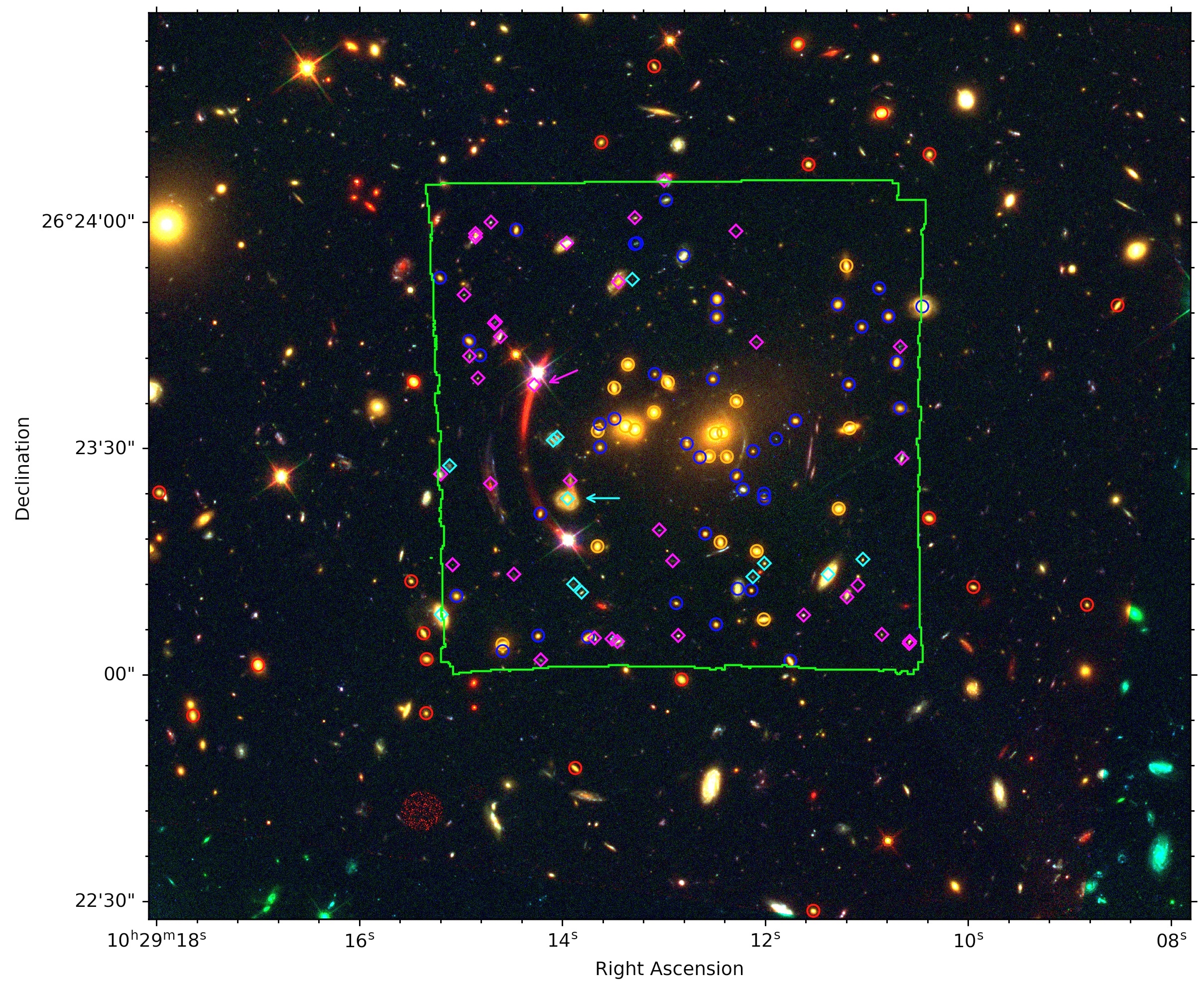} \\
\includegraphics[width=0.65\linewidth]{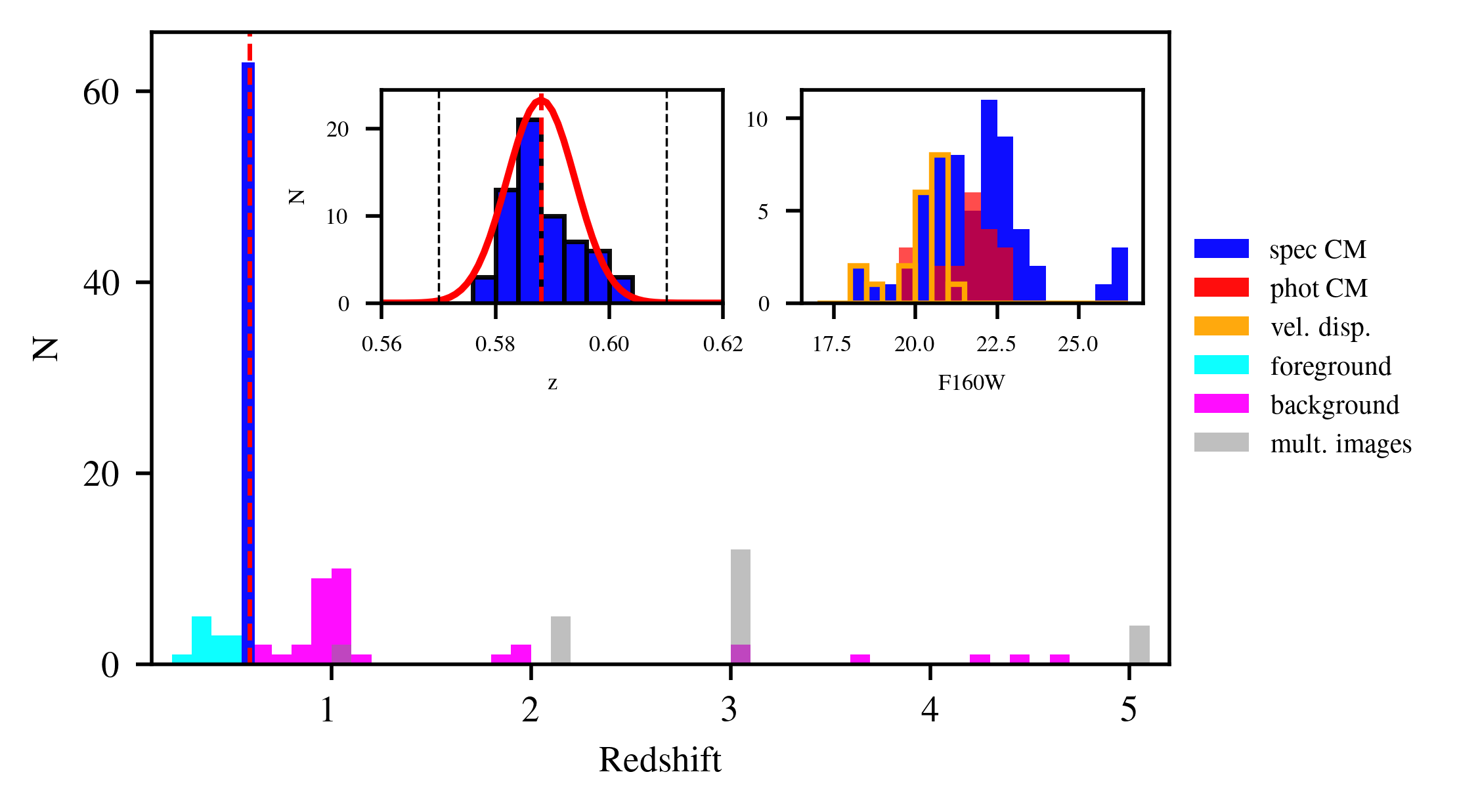}
\caption{\textbf{Top:} Color-composite RGB image of SDSS1029, constructed with the HST/ACS passbands F435W (blue), F814W (green), and F160W (red).
The MUSE footprint is shown in green. We show the spectroscopic and photometric cluster galaxies (as blue and red circles, respectively), and those with a reliable measurement of their stellar velocity dispersion (yellow circles). Foreground and background objects are labeled as cyan and magenta diamonds, respectively. The foreground and background objects highlighted with arrows are included in the SL modeling as described in Sections \ref{sec:FGBG} and \ref{sec:MassMods}.
\textbf{Bottom:} MUSE spectroscopic redshift distribution of the objects identified ($\mathrm{QF \geq 2}$) in the core of SDSS1029. The top left inset shows a zoom-in around the cluster redshift $z = 0.588$, as shown by the mean of the Gaussian (in red). The black dashed vertical lines locate the redshift interval [0.57, 0.61] which includes the 63 MUSE cluster members spectroscopically confirmed (of which 6 objects have a QF = 1). The top right inset shows the distribution of cluster members included in the SL modeling as a function of their magnitudes in the HST/F160W filter, following the same color-coding as in the top image.
} \label{histocm}
\end{figure*}

\section{Observations and Data} \label{sec:data}
This section presents the photometric and spectroscopic data sets used in this work.

\subsection{HST imaging} \label{sec:hst}
We use archival HST multi-color imaging (GO-12195; P.I.: Oguri), from the \textit{Advanced Camera for Surveys} (ACS) and the \textit{Wide Field Camera 3} (WFC3), taken between April and May 2011. SDSS1029  was imaged with ACS/F475W (2 orbits), ACS/F814W (3 orbits) and WFC3/F160W (2 orbits). A description of the observations and data reduction process is detailed in \cite{Oguri2013}. 
We extract the photometric catalogs by using the software \texttt{SExtractor} \citep{Bertin1996} in dual-mode, with the F814W band as the detection image. We apply a two-step extraction technique. In a first step, the bright and extended sources are properly deblended using the \textit{cold mode}, then we set the configuration parameters in \textit{hot mode}, in order to detect faint objects and to properly split close sources \cite[see e.g.,][]{rix04, cal08}. Finally, we use \texttt{GALFIT} \citep[][]{Peng2010} to measure the magnitude values of 20 objects that are affected by the presence of bright close neighbors: 19 cluster galaxies and 1 foreground galaxy (see Tables \ref{tab:FB} and \ref{tab:cm}).

%---------------------------------------------------------------

   \begin{figure}
   \centering   
   \includegraphics[width=0.48\textwidth]{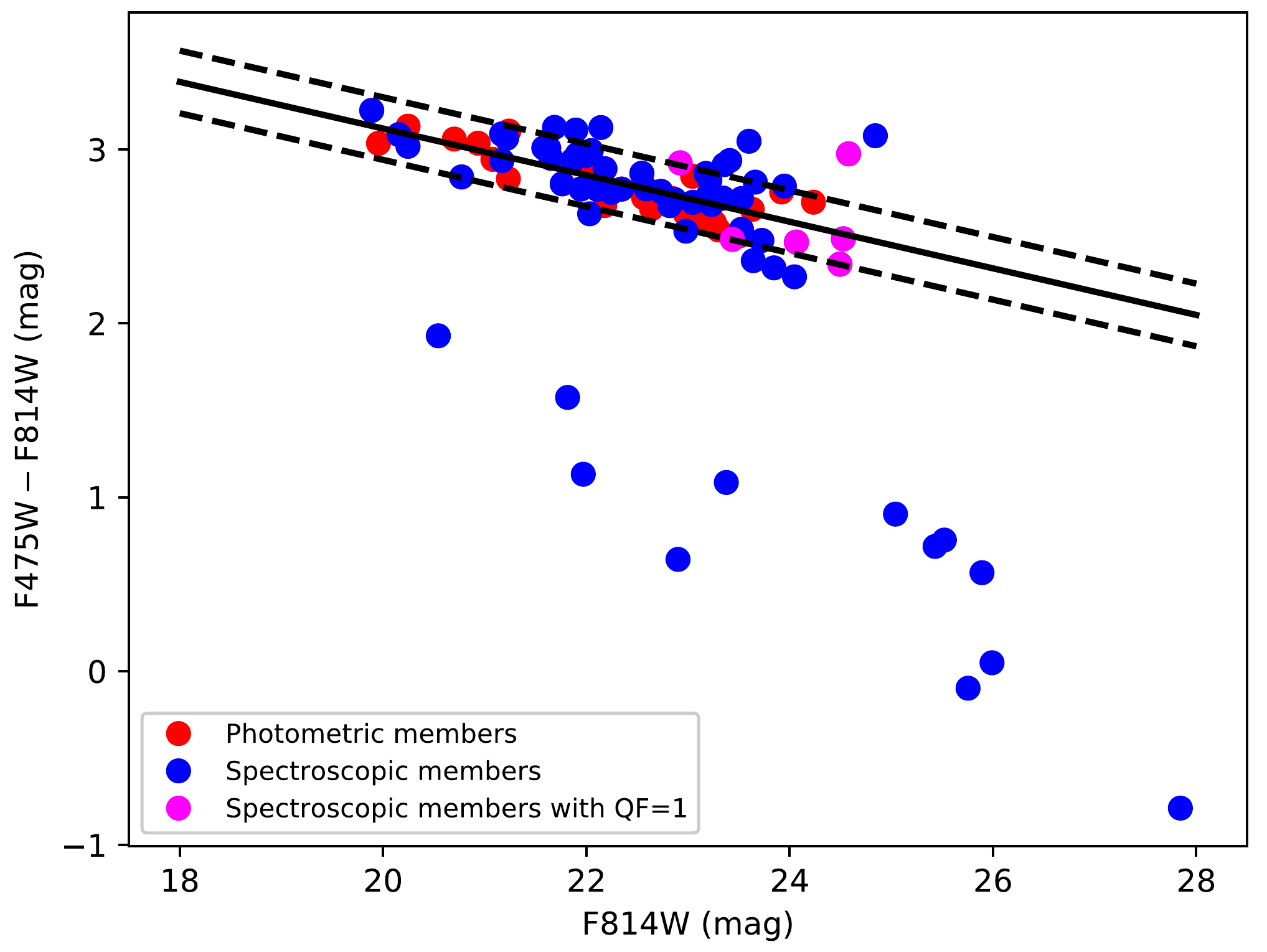}
   \caption{Color-Magnitude relation of SDSS1029 spectroscopic members (blue dots). Red dots mark the cluster members photometrically selected, according to the red-sequence, and the magenta dots are the spectroscopic members with QF = 1. The best-fit of the spectroscopic red sequence is indicated in black and the dashed lines represent the 68\% confidence limits.}
          \label{figCM}%
    \end{figure}   
%---------------------------------------------------------------

%---------------------------------------------------------------
\begin{figure*}
\centering
\includegraphics[width=0.6\linewidth]{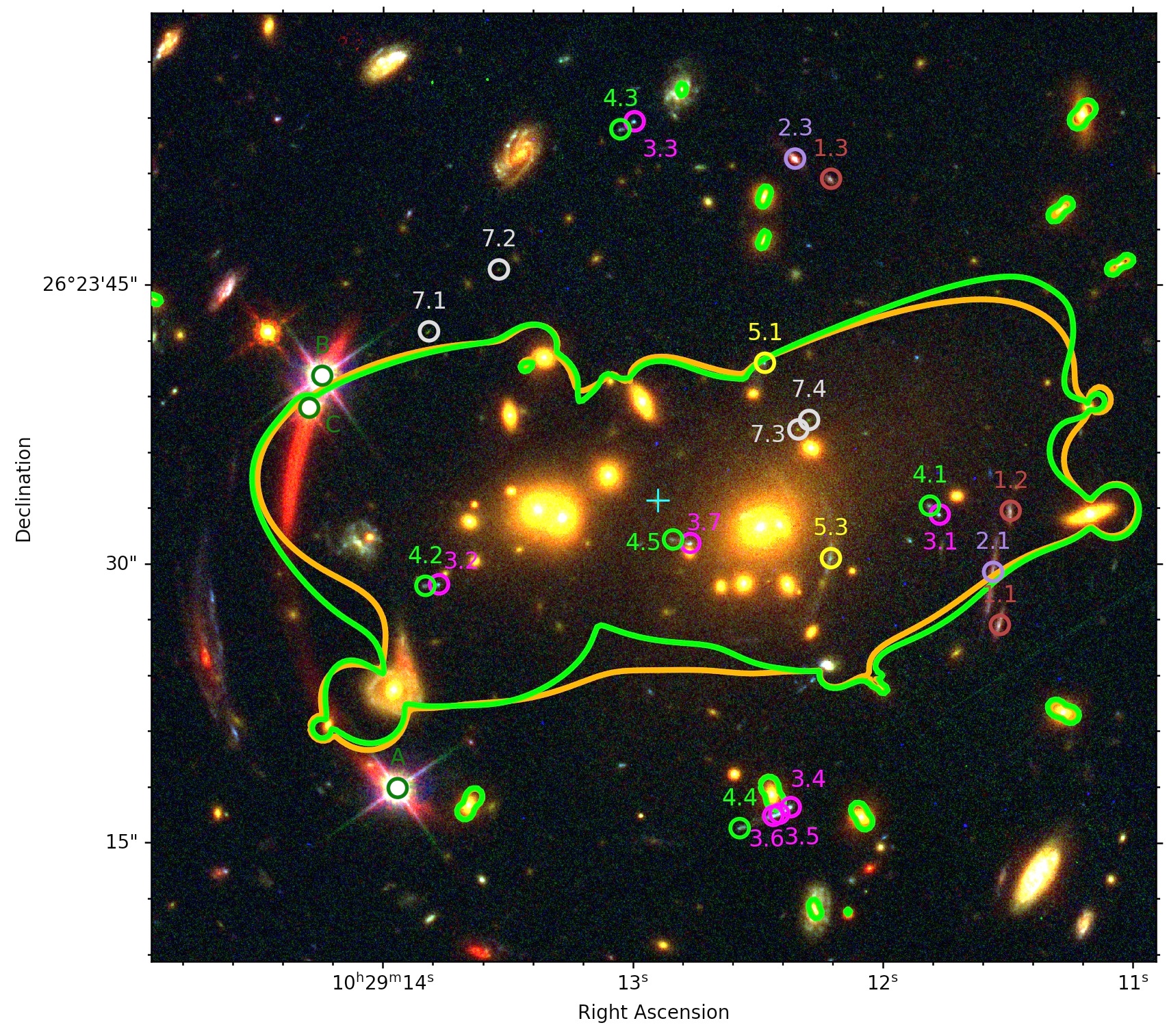}
\caption{RGB image of the central region of SDSS1029, where several multiple images have been identified. We show here the images included in our SL models (listed in Table \ref{tab:multim}), following the same notation as in \citet{Oguri2013}. 
The resulting critical curves at the quasar redshift ($z=2.1992$) from our strong lensing best-fit Model 1 (in orange) and Model 2 (in green) are displayed. The cyan cross indicates the reference position from which the cumulative projected mass and average surface mass density profiles are computed.} 
\label{fig:multim}
\end{figure*}
%---------------------------------------------------------------

\subsection{VLT/MUSE spectroscopy} \label{sec:vlt}
SDSS1029 was recently observed with the integral field spectrograph MUSE, mounted on the VLT, under the program 0102.A-0642(A) (P.I.: C. Grillo). The data were acquired between March 8th, 2019 and May 3rd, 2019. 
Targeting the cluster with a single pointing ($\sim$~1~arcmin$^2$), a total of 12 exposures of 1440 seconds each were obtained, with a cumulative exposure time of $4.8$ hours on target.

We use the standard reduction pipeline \citep[version 2.6,][]{2020A&A...641A..28W} to process the raw exposures and create the final stacked data-cube.
All standard corrections (bias, flat-field, etc.) and calibrations (flux and wavelength) are applied.
Moreover, we use the \emph{auto-calibration} method implemented in the pipeline to mitigate the slice-to-slice flux variations.
Since the sky-subtraction performed by the pipeline is known to leave significant residuals, we also apply the Zurich Atmosphere Purge \citep[ZAP,][]{2016MNRAS.458.3210S}.
During the inspection of each of the 12 exposures, we notice that in one case the guide star tracking failed, strongly degrading the image quality of this specific exposure.
The final data-cube is composed of 11 exposures, totaling 4.4 hour depth on target, has a field of view of $\sim 1\arcmin.1$ across (see the top panel of Figure \ref{histocm}) and covers the wavelength range 4750 \r{A} - 9350 \r{A}, with a resolution of $\sim$ 2.4 \r{A}.
The data has a spatial pixel size of $0\arcsec.2$ and the FWHM measured from the white image is $0\arcsec.71$.

To measure the redshifts of all objects located in the MUSE field of view, we extract 1-dimensional spectra of all sources with HST detections within a circular aperture of $0.\arcsec8$ radius.
For faint galaxies, we use custom apertures based on their estimated morphology from the HST imaging.
With the help of spectral template matching of different galaxies, as well as the identification of emission lines (such as \ion{O}{ii}, Balmer lines, Lyman-$\alpha$, UV carbon lines, etc.), we build our redshift catalog.
To each redshift measurement is then assigned a Quality Flag (QF), which quantifies its reliability \citep[see][]{Balestra2016, Caminha2016}: \textit{insecure} (QF = 1), \textit{likely} (QF = 2), \textit{secure} (QF = 3), and \textit{based on a single emission line} (QF = 9).

The full spectroscopic sample contains 127 reliable (i.e., QF $\ge$ 2) redshift measurements, of which 1 is a star, 12 are foreground ($z < 0.57$) galaxies, 57 are cluster members ($0.57\le z \le 0.61$, see Section \ref{sec:masscomp}), 36 are background galaxies ($z>0.61$) and 21 are multiple images (see Table \ref{tab:multim}). In particular, three high-redshift sources at $z=4.23, 4.42, 4.66$ are identified. We also add 6 cluster members and 3 multiple images with QF = 1 after a visual inspection of their images and spectra.
The foreground and background objects are shown in Figure \ref{histocm} and their coordinates and redshifts are listed in Table \ref{tab:fgbgmuse}. The multiple images and cluster members catalogs are presented in Section \ref{sec:SLM}.

\section{Strong Lensing Modeling} \label{sec:SLM}
We model the total mass distribution of SDSS1029 with the public software \texttt{lenstool}\footnote{\url{https://projets.lam.fr/projects/lenstool}} which adopts a parametric mass reconstruction \citep{Kneib1996, Jullo2007}. In this section, we present the multiple images used to optimize the model, the cluster member selection and a brief summary of the methodology. We refer to the publications mentioned above for further details \citep[see e.g.][]{Caminha2016}.

\subsection{Cluster member selection}  \label{sec:masscomp}
The cluster galaxies that are included in the strong lensing model are selected based on both spectroscopic and photometric information. 

Firstly, we identify 57 galaxies with a reliable redshift estimate (i.e. with a QF $\geq$ 2). Figure~\ref{histocm} shows the redshift distribution of these galaxies, which can be fit with a Gaussian distribution with mean and standard deviation values of $\overline{z}=0.588\pm0.006$. We also identify 6 additional galaxies in the central region of the cluster with an uncertain redshift estimate (QF = 1), but bright enough (F160W $<23.5$) to introduce significant perturbations to the overall lens model. These spectroscopic cluster members are selected as the galaxies with rest-frame relative velocities within $\Delta V = 3000 ~ \rm km\,s^{-1}$ from the cluster mean velocity, which corresponds to the redshift range $z=0.57$-$0.61$.
We label as the two brightest cluster galaxies (BCGs) the two central galaxies (IDs 1933 and 1953 in Table \ref{tab:cm}). Each BCG has a close neighbor (within a projected distance of $\sim 1\arcsec.5$), which are identified in Table \ref{tab:cm} as IDs 1964 and 1849, respectively.
We complete the spectroscopic sample by selecting 20 additional photometric members lying on the cluster red sequence (see the red dots in Figure~\ref{figCM}) and having F160W $\le$ 24. 
To determine the cluster red sequence, we fit the color-magnitude relation by considering all the spectroscopically confirmed member galaxies. The fit to the relation is obtained using the least-trimmed squares (LTS) technique implemented by \citet{Cappellari2013}. The black solid line in Figure \ref{figCM} shows the red sequence, while the dashed lines indicate the 1$\sigma$ measured scatter of 0.18. 

Our final cluster member catalog, including 83 cluster member galaxies which are integrated in the following lensing analysis, is presented in Figure \ref{histocm}, together with the galaxy redshift distribution and F160W magnitudes. The complete catalog is also given in Appendix \ref{sec:A3}, Table~\ref{tab:cm}.

\subsection{Multiple image systems}  \label{sec:multim}
In this work, we consider both previous identifications of multiple image systems, available in the literature \citep{Oguri2013}, and a new system, discovered from an inspection of the HST images and the MUSE data-cube. 

The first system lensed by SDSS1029 to be identified was the triply-imaged quasar \citep{Inada2006, Oguri2008}, for which we provide an updated spectroscopic redshift of $z=2.1992$ based on the MUSE data.
We also provide the first spectroscopic confirmation of the remaining 5 lensed systems identified in \cite{Oguri2013}, whose redshift values were originally optimized in the SL modeling. We note that the spectroscopic redshifts of the 5 systems are consistent with the corresponding $2\sigma$ redshift intervals predicted by the strong lensing model presented in \cite{Oguri2013}.
Moreover, we identify the new multiply imaged source 7, at a redshift $z=5.0622$. The number and positions of four multiple images are well reproduced by our lens models. In addition to these confirmed four images, an extra, fainter image is predicted outside the MUSE field-of-view, in the southern region of the cluster.
In our models, we do decide not to consider families or images identified in \cite{Oguri2013} that do not a have a clear emission peak in the HST photometry, namely system 6 and the individual images 2.2 and 5.2, since their positions cannot be estimated reliably.

All redshift measurements are classified as secure (QF~=~3), but for image 2.1 which is classified as likely (QF~=~2) and the three images of system 1 whose redshift estimate is considered unreliable (QF $<$ 2).
When all redshift measurements of a family have the same quality flag, we adopt the mean value of these measurements as the system redshift. On the other hand, as is the case for system 2, if different quality flags are available, we consider the most reliable one (i.e. a QF = 3) as the redshift of the system.
In order to avoid biases on the determination of the cluster total mass, we consider in our lens models only multiple images with a secure spectroscopic confirmation \citep[see e.g.][]{Grillo2015, Caminha2016, Johnson2016, Cerny2018, Remolina2018}. In the case of system 1, its redshift value is included as a free parameter, optimized during the lensing analysis, and the best-fit redshift value is found to be in agreement with the tentative spectroscopic measurement (see Section \ref{sec:results}).

As all images have a HST detection, we carefully define the central position of each image as the luminosity peak in the HST F814W band. 

The complete sample of multiple image systems spans a large redshift range, between $z=1.02$ and $z=5.06$, with a total of 26 multiple images from 7 background sources. We use the observed positions of the multiple images as constraints in the lens models, providing in total 52 observables and 14 free parameters for the positions of the corresponding sources.
We show the positions of the 26 multiple images in Figure \ref{fig:multim} and their properties are summarized in Table \ref{tab:multim}. The extracted spectra for the objects with a QF $\geq$ 2, with small cutouts of the HST color-composite image, are presented in Figure \ref{specmi}.

\begin{deluxetable}{cccccc}
\tablenum{1}
\tablecaption{Coordinates and spectroscopic redshifts, with the corresponding quality flag, of the multiple image systems.\label{tab:multim}}
\tablewidth{0pt}
\tablehead{
\colhead{ID\tablenotemark{a}} & \colhead{R.A.} & \colhead{Decl} & \colhead{$\rm z_{spec}$}  & QF &\colhead{ID}  \\
\colhead{} & \colhead{(deg)} & \colhead{(deg)} & & \colhead{}  & \colhead{}
}
\startdata
A & 157.3081015 & 26.3883036 & 2.1992 & 3 & 1485\\
B & 157.3093576 & 26.3944634 & 2.1992 & 3 & 1713\\
C & 157.3095761 & 26.3939843 & 2.1992 & 3 & 1789\\
1.1 & 157.2980611 & 26.3907404 & 2.1778 & 1 & 2025\\
1.2 & 157.2978817 & 26.3924467 & 2.1778 & 1 & 2234\\
1.3 & 157.3008758 & 26.3974054 & 2.1778 & 1 & 2786\\
2.1 & 157.2981743 & 26.3915325 & 2.1812 & 2 & 1889\\
2.3 & 157.3014749 & 26.3977063 & 2.1812 & 3 & 2814 \\
3.1 & 157.2990642 & 26.3923892 & 3.0275 & 3 & 2268\\
3.2 & 157.3074114 & 26.3913469 & 3.0275 & 3 & 2136\\
3.3 & 157.3041512 & 26.3982630 & 3.0275 & 3 & 2894\\
3.4 & 157.3015481 & 26.3880193 & 3.0275 & 3 & 1745\\
3.5 & 157.3017377 & 26.3879213 & 3.0275 & 3 & 1733\\
3.6 & 157.3018385 & 26.3878900 & 3.0275 & - & -\\
3.7 & 157.3032208 & 26.3919632 & 3.0275 & 3 & 99992\\
4.1 & 157.2992278 & 26.3925219 & 3.0278 & 3 & 2286\\
4.2 & 157.3076382 & 26.3913247 & 3.0278 & 3 & 99994\\
4.3 & 157.3043869 & 26.3981437 & 3.0278 & 3 & 2877\\
4.4 & 157.3023985 & 26.3877048 & 3.0278 & - & - \\
4.5 & 157.3035100 & 26.3920169 & 3.0278 & 3 & 99993\\
5.1 & 157.3019777 & 26.3946563 & 1.0232 & 3 & 2504\\
5.3 & 157.3008781 & 26.3917377 & 1.0232 & 3 & 2175\\
7.1 & 157.3075794 & 26.3951262 & 5.0622 & 3 & 99998\\
7.2 & 157.3064130 & 26.3960500 & 5.0622 & 3 & 99999\\
7.3 & 157.3014210 & 26.3936610 & 5.0622 & 3 & 999910\\
7.4 & 157.3012420 & 26.3938020 & 5.0622 & 3 & 999911\\
\enddata
%\tablecomments{
\tablenotetext{a}{IDs are presented following the same notation as in \cite{Oguri2013}.}
%}
\end{deluxetable}

\begin{deluxetable}{cccccc}
\tablenum{2}
\tablecaption{Foreground and background galaxies included in the strong lens modeling of SDSS1029. \label{tab:FB}}
\tablewidth{0pt}
\tablehead{
\colhead{ID} & \colhead{R.A.\tablenotemark{a}} & \colhead{Decl\tablenotemark{a}} & \colhead{F160W\tablenotemark{a}} &\colhead{$z_{\rm spec}$} &\colhead{QF}  \\
\colhead{} & \colhead{(deg)} & \colhead{(deg)} & \colhead{(mag)}  & \colhead{} & \colhead{}
}
\startdata
1625\tablenotemark{b} & 157.308162 & 26.389775 & $17.03 \pm 0.10$ & 0.5111 & 3\\
999913\tablenotemark{c} & 157.309500 & 26.393997 & -\tablenotemark{d} & 0.6735 & 2\\
\enddata
\tablenotetext{a}{Coordinates and F160W magnitude (and associated error) are measured with \texttt{Galfit}.}
\tablenotetext{b}{This galaxy is also referred to as FG in the text.}
\tablenotetext{c}{This galaxy is also referred to as GX in the text.}
\tablenotetext{d}{Being very close to the image C of the quasar, a reliable magnitude measurement could not be obtained.}
\end{deluxetable}

\subsection{Modeling methodology} \label{sec:method}
We model the total mass distribution of SDSS1029 as the sum of several components: a diffuse mass component, representing the contribution of the dark matter and intra-cluster medium, together with small-scale halos, that account for the cluster galaxies (baryons and underlying dark matter).

The diffuse mass distribution is parametrized as a dual pseudo-isothermal elliptical mass profile \citep[dPIE,][]{Eliasdottir2007}. The free parameters associated with this profile are the following: the coordinates, $x,~y$;  the ellipticity defined as $\epsilon=(a^2-b^2)/(a^2+b^2)$, where $a$ and $b$ are the values of the major and minor semi-axes, respectively; the orientation, $\theta$ (counted counter-clockwise from the $x$-axis); the core and cut radii, $r_{\rm core}$ and $r_{\rm cut}$; and a velocity dispersion, $\sigma_{\rm LT}$, which is related to the central velocity dispersion of the dPIE profile as $\sigma_0 = \sqrt{3/2}~ \sigma_{\rm LT}$.
The associated 3D density distribution, as presented in \cite{Limousin2005}, is defined as:
\begin{equation}
\rho(r)=\frac{\rho_0}{(1+r^2/r_{\rm core}^2)(1+r^2/r_{\rm cut}^2)},
\end{equation}
where $r$ is the distance from the center, and $r_{\rm core}$ and $r_{\rm cut}$ are the core radius and truncation radius, respectively.
The dPIE density profile is characterized by two changes in the slope: within the transition region ($r_{\rm core} < r < r_{\rm cut}$), it behaves as an isothermal profile, where $\rho \propto r^{-2}$, while at larger radii, the profile falls more steeply, as $\rho \propto r^{-4}$.

On smaller scales, each halo associated with a cluster galaxy is modeled as a spherical dPIE with a vanishing core radius. 
In order to significantly reduce the number of free parameters in the modeling, the following scaling relations between the sub-halo total mass and its associated luminosity \citep{Jullo2007} are adopted:

\begin{equation}
\label{eqsigma}
\sigma_0=\sigma_0^{\star} \left(\frac{L}{L^{\star}}\right)^{\alpha},
\end{equation}
\begin{equation}
\label{eqrcut}
r_{\rm cut}=r_{\rm cut}^{\star}\left(\frac{L}{L^{\star}}\right)^{\beta},
\end{equation}
where $L^{\star}$ is the reference luminosity of a galaxy at the cluster redshift, that we associate with the brightest cluster galaxy ($\mathrm{F160W=18.24}$), $\sigma_0^{\star}$ and $r_{\rm cut}^{\star}$ are the two parameters optimized in the lens analysis, and $\alpha$ and $\beta$ represent the slopes of the $\sigma_0$ and $r_{\rm cut}$ scaling relations, respectively.
As detailed in \citet{Bergamini2019, Bergamini2021} (hereafter B19 and B21, respectively), we can define the total mass-to-light ratio as $M^{\rm tot}_i/L_i \propto L ^{\gamma}_i$, where the relation between the slope parameters can be expressed as $\beta =\gamma-2\alpha+1$. If $\gamma = 0.2$, the scaling relations are consistent with the so-called tilt of the fundamental plane \citep{Faber1987, Bender1992}.

Once the mass components are defined, described by a set of model parameters $\mathbf{p}$, the best-fitting values of the model parameters are found by minimizing on the image plane the distance between the observed, $\boldsymbol{\theta}^{\rm obs}$, and model-predicted, $\boldsymbol{\theta}^{\rm pred}$, positions of the multiple images. 
This is done by minimizing the $\chi^2$ function, defined on the image plane as follows:
\begin{equation}
\chi^2(\mathbf{p})=\sum_{j=1}^{N_{\rm fam}}\sum_{i=1}^{N_{\rm img}^{j}}\left(\frac{\left|\boldsymbol{\theta}^{\rm obs}_{ij}-{\boldsymbol{\theta}}^{\rm pred}_{ij}(\mathbf{p})\right|}{\sigma_{ij}}\right)^2, 
\end{equation}
where $N_{\rm fam}$ and $N_{\rm img}^{j}$ are the total number of families and images for the family $j$ included in the modeling, respectively. The positional uncertainty for the observed images is $\sigma_{ij}$.
We adopt an initial positional uncertainty of $0\arcsec.25$ for most images, but allow for a larger uncertainty, of $0\arcsec.5$, for the multiple images of families 2 and 5, as their exact positions are less reliably constrained. For the final model, the positional uncertainty is later rescaled in order to have a minimum $\chi^2$ value comparable with the number of degrees of freedom ($\nu$), i.e. to get $\chi^2/\nu\sim1$.

%--------------------------------------------------------------------

   \begin{figure}
   \centering   
   \includegraphics[width=0.48\textwidth]{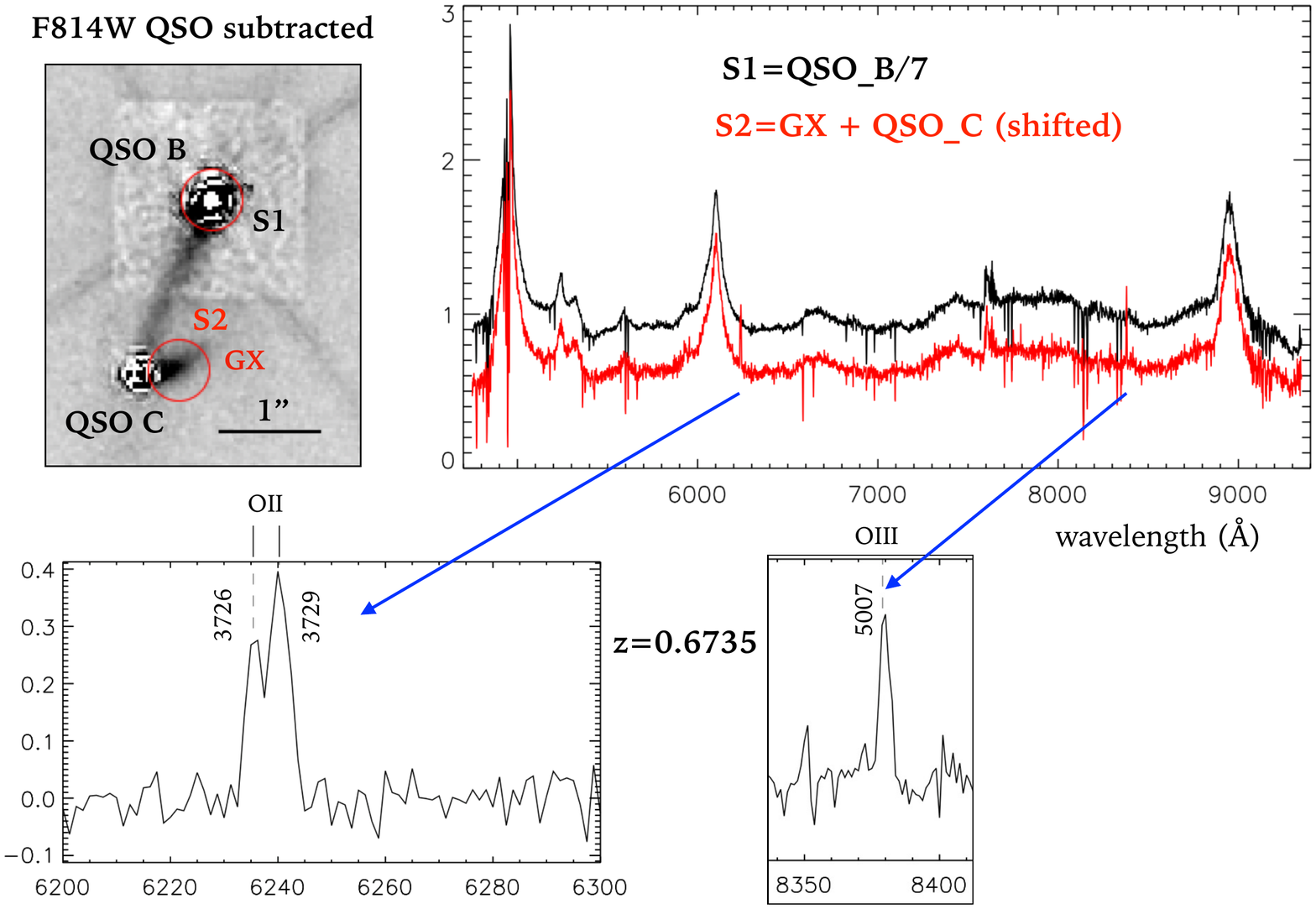}
   \caption{\textbf{Top:} MUSE 1D spectra of the quasar (QSO) image B (S1 in black, rescaled by a factor of 7) and the hidden galaxy GX lying close to it, highly contaminated by the QSO image C (S2 in red). Both spectra are extracted from circular apertures of 0\arcsec.3 radius, indicated in red in the HST F814W image cutout, where QSO B and C have been subtracted.  \textbf{Bottom:} Two portions of the subtracted 1D spectrum (S2$-$S1), where the [\ion{O}{ii}] and [\ion{O}{iii}] emission lines of a source at a redshift of 0.6735 are highlighted.}
          \label{fig:hiddengal}
    \end{figure}   
%--------------------------------------------------------------------

Other statistical estimators are also used to assess the goodness of our reconstructions. In particular, the root mean square (rms) value of the difference between the observed and model-predicted positions of the multiple images,
\begin{equation}
{\rm rms}=\sqrt{\frac{1}{N_{\rm tot}}\sum_{i=1}^{N_{\rm tot}}\left|\boldsymbol{\theta}^{\rm obs}_{ij}-{\boldsymbol{\theta}}^{\rm pred}_{ij}(\mathbf{p})\right|^2},
\end{equation}
where $N_{\rm tot}$ is the total number of images.
Finally, we consider the Bayesian information criterion \citep[BIC,][]{Schwarz1978} and the corrected Akaike information criterion \citep[AICc,][]{Akaike1974}, which is suitable for models with a relatively low number of constraints.
They are defined as follows:
\begin{equation}
{\rm BIC}=-2\ln({\mathcal{L}}) + k \times \ln(n),
\end{equation}
\begin{equation}
{\rm AICc}=2k -2\ln({\mathcal{L}}) + \frac{2k(k+1)}{(n-k-1)},
\end{equation}
where $\mathcal{L}$ is the maximun value of the likelihood (see \citealt{Jullo2007}), $k$ is the number of free parameters and $n$ is the number of constraints. 

\begin{figure}
\centering
\includegraphics[width=\columnwidth]{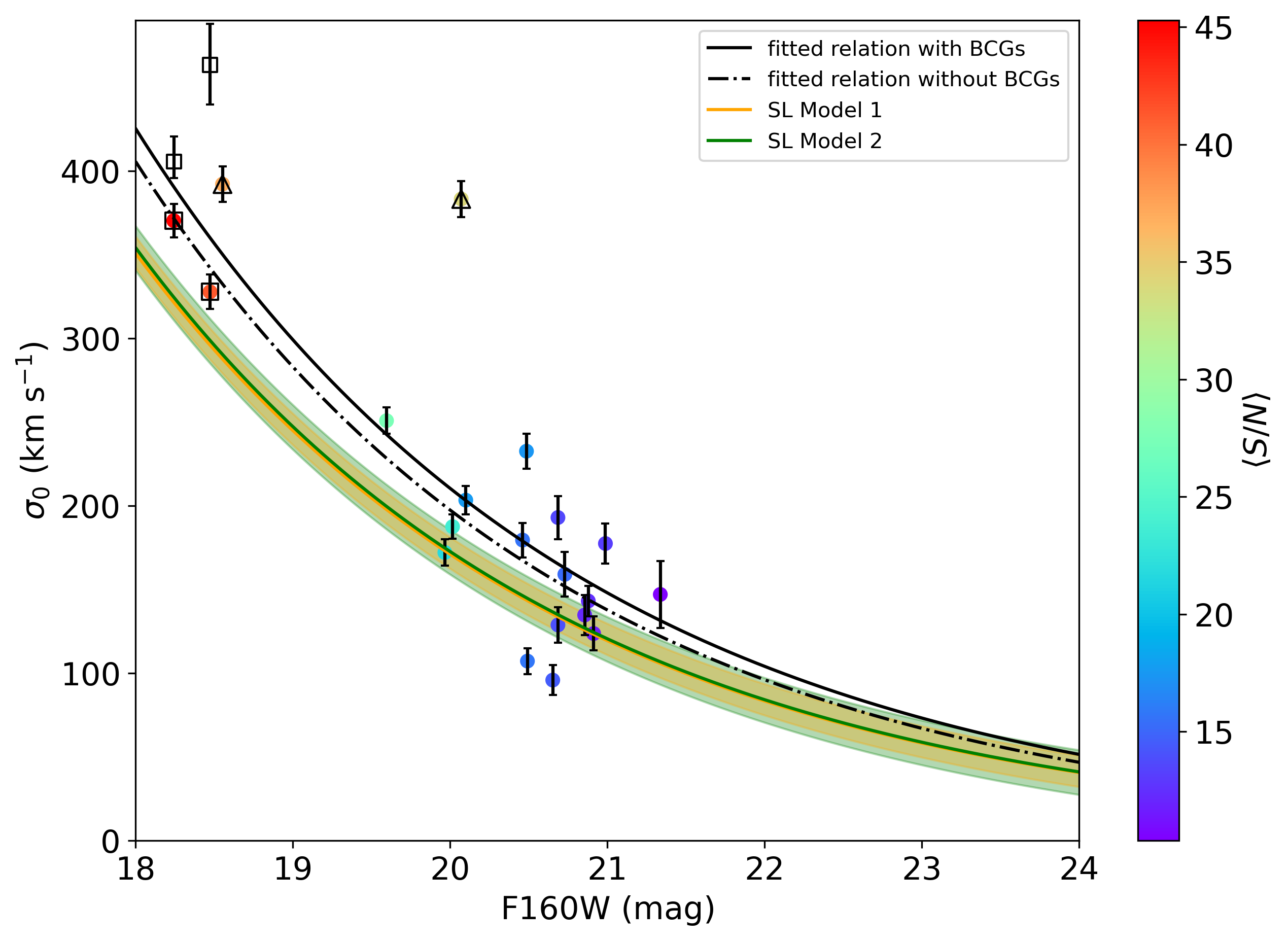} 
\caption{Measured stellar velocity dispersions of a sub-sample of 20 cluster members as a function of their magnitudes in the HST F160W filter (filled circles, whose colors depend on the $\langle S/N\rangle$ of their spectra).
The black filled squares and triangles indicate the measurements of the 4 central galaxies. 
The black solid and dashed lines are the best-fit
$\sigma_0$-F160W relations, obtained as described in Section \ref{sec:veldisp}. The orange and green lines and areas correspond to the median and the 68\% confidence level of the $\sigma_0$-F160W relation from our SL Models 1 and 2, respectively. The resulting relations from the two lens models are consistent and do, in fact, overlap. We also mark as black empty squares the median velocity dispersions and associated $1\sigma$ uncertainties for the two BCGs which are individually optimized in Model~2.
} \label{fig:VELDIS}
\end{figure}

\subsection{A bright foreground and a hidden galaxy}  \label{sec:FGBG}
This section provides a description of two galaxies, one in the foreground and one in the background of the cluster, that are included in the strong lens modeling, because of their vicinity in projection to the images A and C of the quasar. The properties of both objects are summarized in Table \ref{tab:FB}.

The foreground ($z=0.5111$) object is a bright spiral galaxy located $\sim 5\arcsec.3$ apart from the southern image A of the quasar (see Figure \ref{fig:multim}). 
We also include the galaxy that sits extremely close, at $\sim0\arcsec.4$, to the image C of the quasar and that was first identified in the HST images by \cite{Oguri2013}, since it is known that substructures can affect the model-predicted values of time delays \citep{Fohlmeister2007} and magnifications. Indeed, \citet{Oguri2013} showed that this galaxy, previously labeled as GX and considered in the lens model (at the cluster redshift), had a significant impact on the derived magnification of the quasar images, resolving the observed ``flux ratio anomalies".\\
Following an inspection of the MUSE data-cube, we extract the spectrum of this object within a aperture of $0\arcsec.3$ radius, as shown in Figure \ref{fig:hiddengal}. The obtained spectrum (S2) is affected by the contribution of both image C of the quasar and the perturber GX. We then subtract to it a properly re-scaled spectrum of image B of the quasar (S1, which is not contaminated). The subtracted spectrum (S2$-$S1) clearly shows the [\ion{O}{ii}] $ \lambda \lambda ~ 3727,3729$ \r{A} doublet and the [\ion{O}{iii}] $\lambda \lambda ~ 5007$ \r{A} emission lines. We can then reliably measure a redshift of $z=0.6735$, allowing us to confirm the background nature of the galaxy GX.

\begin{deluxetable*}{lrcccccccccc}
	\tablenum{3}
	\tablecaption{Strong lensing mass model parameters for Model 1 (top) and Model 2 (bottom). Median values and 68\% confidence level intervals are quoted. Parameter values in square brackets are kept fixed. \label{tab:BFmod1}}
	\tablewidth{0pt}
	\tablehead{
	\colhead{Model Statistics} & \colhead{Component} & \colhead{$x$} & \colhead{$y$} & \colhead{$\varepsilon$} & \colhead{$\theta$} & \colhead{$\sigma_0$} & \colhead{r$_{\rm cut}$} & \colhead{r$_{\rm core}$} & \colhead{$\gamma_{\rm ext}$} & \colhead{$\phi$}\\
	\colhead{}	&  \colhead{} & \colhead{[\arcsec]} & \colhead{[\arcsec]} &  \colhead{} & \colhead{[$\deg$]} & \colhead{[km\ s$^{-1}$]} & \colhead{[\arcsec]} & \colhead{[\arcsec]} & \colhead{} & \colhead{[$\deg$]}
}
	\startdata
	rms = 0\arcsec.15\ & DM\_1953 & $-12.5^{+0.6}_{-0.5}$ & $2.8^{+0.2}_{-0.3}$ & $0.63^{+0.05}_{-0.06}$ & $1.6^{+3.7}_{-2.1}$ & $594^{+72}_{-28}$& [2000] & $2.4^{+1.8}_{-1.0}$& -- & --\\ 
	$\chi^2/\nu$ = 0.71 &  DM\_1933 & $2.2^{+0.5}_{-0.4}$ & $0.3^{+0.4}_{-0.3}$ & $0.66^{+0.04}_{-0.06}$ & $36.3^{+1.8}_{-1.7}$ &$ 763^{+26}_{-39}$ & [2000] & $6.0^{+0.3}_{-0.5}$ & -- & --\\ 
	$\log$($\mathcal{L}$) = $31.33$ &  1953 & [$-12.0$] & [$1.0$] &  -- &  --& -- & --  & -- & -- & --\\  
    $\log$($\mathcal{E}$) = $-43.95$  &  1933 & [0] & [0] &  -- &  --& -- & --  & -- & -- & --\\  
    BIC = 75.63  &FG & [$-19.7$] & [$-8.8$] & [0]  & [0]  & $242^{+18}_{-18}$ & $20.0^{+0.2}_{-14.3}$  & [0]  & -- & --\\ 
    AIC = 7.34 & GX & [$-24.0$] & [$6.4$] & [0]  & [0]  &  $58^{+24}_{-27}$ & $ 9.1^{+12.8}_{-0.3}$& [0] & -- & --\\ 
    AICc = 164.84 & $L^{*}$ Galaxy & -- & -- & -- & -- & $322^{+10}_{-9}$ & $17.5^{+7.1}_{-4.1}$ & [0] & -- & --\\ 
    & Ext. Shear & -- & -- & -- & -- & -- & -- & -- & $0.09^{+0.02}_{-0.02}$ & $65.9^{+4.4}_{-8.1}$\\
    \hline
    rms = 0\arcsec.22\ & DM\_1953 & $-11.8^{+2.1}_{-4.0}$ & 0.4$^{+0.9}_{-1.2}$ & 0.78$^{+0.06}_{-0.08}$ & $-4.6^{+2.9}_{-3.2}$ & 763$^{+40}_{-54}$ & [2000] & 12.9$^{+3.1}_{-1.9}$ & -- & --\\ 
	$\chi^2/\nu$ = 1.03 &  DM\_1933 & 5.5$^{+0.6}_{-1.2}$ & 4.1$^{+1.0}_{-0.9}$ & 0.63$^{+0.06}_{-0.06}$ & 32.6$^{+4.0}_{-3.5}$ & 653$^{+49}_{-45}$ & [2000] & 5.9$^{+0.8}_{-0.8}$ & -- & --\\ 
	$\log$($\mathcal{L}$) =  15.87 & 1953 & [$-12.0$] & [$1.0$] & [0] & [0] & 463$^{+25}_{-24}$ & 25.5$^{+16.9}_{-9.0}$  &  [0] & -- & --\\ 
	$\log$($\mathcal{E}$) = $-33.99$  & 1933 &[0] & [0]& [0] & [0] & 406$^{+10}_{-15}$ & 16.5$^{+2.4}_{-6.9}$  &  [0] & -- & --\\ 
	BIC = 114.46 & FG & [$-19.7$] & [$-8.8$] & [0]  & [0]  & 197$^{+27}_{-48}$  & 17.6$^{+4.5}_{-13.7}$  & [0]  & -- & --\\ 
	AIC= 42.26 & GX & [$-24.0$] & [$6.4$] & [0]  & [0]  &  78$^{+23}_{-22}$ & 9.7$^{+12.9}_{-0.4}$ & [0] & -- & --\\ 
	AICc=  243.12 & $L^{*}$ Galaxy & -- & -- & -- & -- & 325$^{+13}_{-14}$ & 20.7$^{+11.3}_{-6.5}$ & [0] & -- & --\\ 
	\enddata
\end{deluxetable*}

\subsection{Internal kinematics of cluster members}  \label{sec:veldisp}

Extensive spectroscopic campaigns have allowed for a large number of multiple images to be identified, with measured spectroscopic redshifts. However, lens models still suffer from some degeneracies between the different mass components \citep{Limousin2016}.
Indeed, differences between the scaling relations predicted by lens models without prior kinematic information and the measured line-of-sight stellar velocity dispersion for the cluster members have been previously reported for some well studied clusters (B19).
Such degeneracies can be however reduced when using additional independent information.

In this work, we use stellar kinematics from a subset of cluster members to better constrain the sub-halo population of the cluster.
We follow the same methodology as in B19 and B21. A detailed description of the methodology, and comparison with simulations is given in these references, but we provide here a brief overview. 

Spectra of all cluster members are extracted from the MUSE data cube within $0\arcsec.8$ radius apertures, which has shown to provide the best compromise between a high $\langle S/N \rangle$ value and a low contamination from the intra-cluster-light (ICL) or angularly close sources (the median FWHM value of these MUSE observations is $0\arcsec.71$, see Section \ref{sec:vlt}). All the extracted spectra are then visually inspected, and if the contamination from the ICL is significant, the velocity dispersion of a faint galaxy is discarded, as it is likely biased high.
To measure the line-of-sight stellar velocity dispersions, we use the public software Penalized Pixel-Fitting method \citep[pPXF,][]{Cappellari2004, Cappellari2017}, and cross-correlate the observed spectra with an extended set of stellar templates in the rest-frame wavelength range [3600, 5000] \r{A}. 

In order to only consider reliable line-of-sight velocity dispersion measurements, we follow the prescription adopted in B19 and B21, and we limit the sample to those galaxies with $\langle S/N \rangle >10$ and $\sigma_0 > 80~ \mathrm{km~s^{-1}}$.
The final sample of cluster members with measured velocity dispersions includes 20 objects down to $ \mathrm{F160W} \sim 21$ (see Figure \ref{histocm}). We also show in
Figure \ref{fig:VELDIS} the measured velocity dispersion values as a function of the F160W magnitudes.

Following B19 and B21, we adopt a Bayesian approach to fit the measured velocity dispersions of the 20 cluster members. We can then derive the best-fit value of the slope $\alpha$ and the reference $\sigma_{0}^{\star}$ of the $\sigma_0$-F160W relation.
We note, however, that as the four central cluster galaxies of SDSS1029 (the two BGCs and two close neighbors, see Section \ref{sec:SLM}) lie very close together (within $1\arcsec.5$, see Figures \ref{histocm} and \ref{fig:multim}), their extracted spectra, from which the values of the stellar velocity dispersions are then measured, can be contaminated by their close neighbors, artificially increasing our estimates. As illustrated in Figure \ref{fig:VELDIS}, we perform two fits, including or not, these four central galaxies. We find that the resulting fits are very similar, with $\Delta \alpha \sim0.01$ and $\Delta \sigma_{0} \sim 20 \mathrm{\, km \, s^{-1}}$, the latter difference being included in the adopted measurement uncertainty. We choose to adopt in the following analysis the fit not including the four central galaxies.
The posterior of the Bayesian fit is in turn used as a prior for the scaling relations in our strong lens models (see Section \ref{sec:MassMods}). 

\begin{figure}
\centering
\includegraphics[width=\columnwidth]{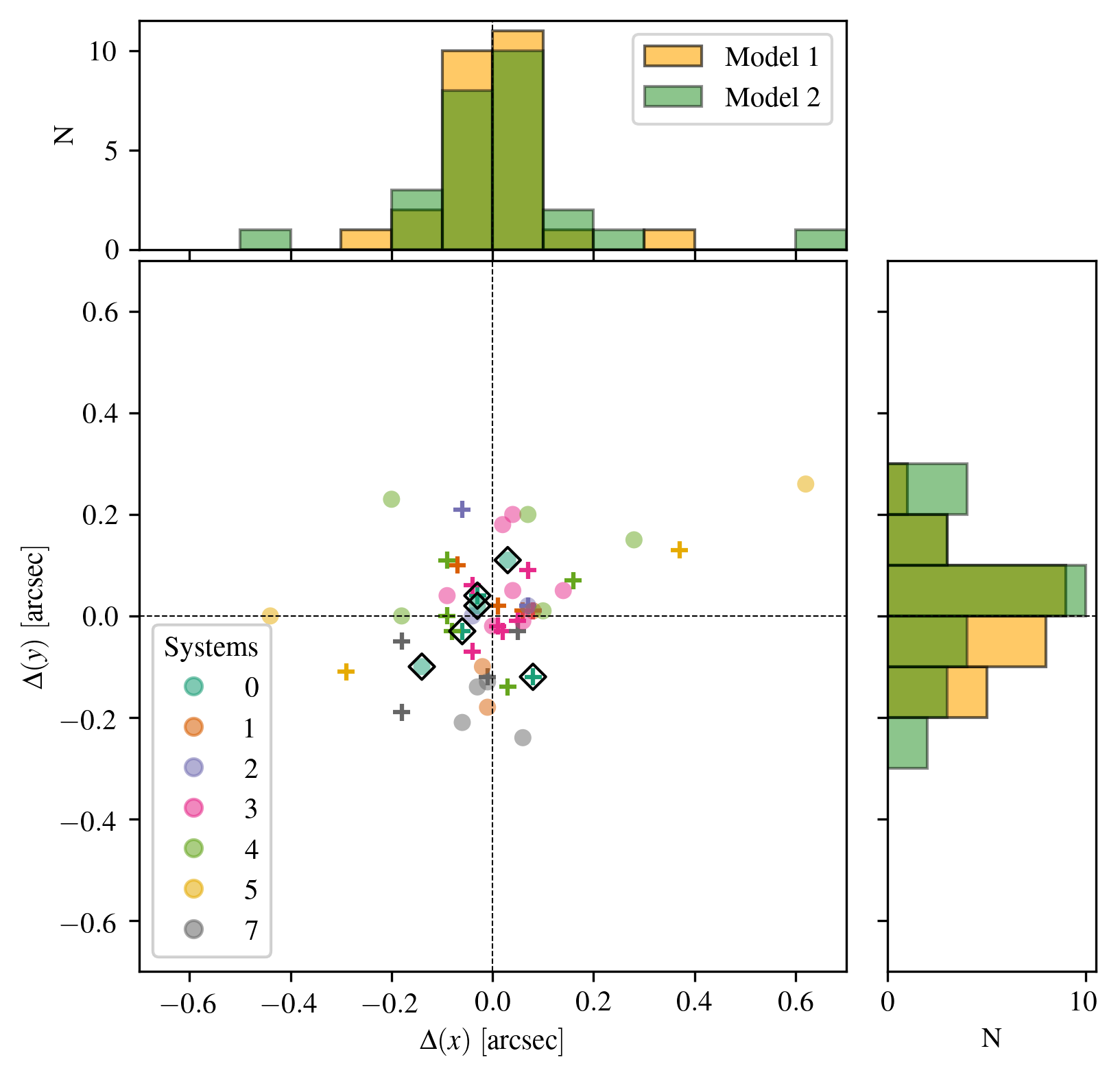} 
\caption{The middle panel shows the distribution of the separations $\Delta (x)$ and $\Delta (y)$ between the observed and model-predicted image positions as crosses(circles) for Model 1(2). Each system is color-coded as in Figure \ref{fig:multim}. The separations for the quasar images are, in addition, highlighted with black diamonds. The upper and right panels show the marginalized distributions of the $\Delta (x)$ and $\Delta (y)$ separations, respectively.
} \label{fig:dxdy}
\end{figure}

\subsection{Mass models} \label{sec:MassMods}
We consider several mass models in order to investigate which parametrization best reproduces the observed multiple image positions. We describe here two different and complementary SL models, \textit{Model 1} and \textit{Model 2}, hereafter. Both models adopt the same catalog of multiple images (with the same initial positional uncertainties) and cluster members, as presented in Sections \ref{sec:multim} and \ref{sec:masscomp}. The redshift of system 1 is also optimized in both models, assuming a large flat prior. 

Regarding the mass parametrization, both model include two large-scale DM halos, whose central coordinates are free to vary within $\pm7\arcsec$ around the corresponding BCG positions. While the cut radius is fixed to a very large value, the ellipticity, position angle, core radius and velocity dispersion are optimized within large flat priors.
Cluster members are modeled within the scaling relations. Following the stellar velocity dispersion measurements for the sub-sample of cluster members presented in Section \ref{sec:veldisp}, we use a Gaussian prior for the value of the normalization $\sigma_{0}^{\star}$ centered on the measured value of $371~\rm km~s^{-1}$, with a standard deviation of $91 \mathrm{\, km \, s^{-1}}$.
A large prior for $r_{\rm cut}^{\star}$ is adopted. 
We also fix the value of the slope $\alpha$ to the fitted value of 0.39 and we infer the value of $\beta$ so that $M^{\rm tot}_i/L_i \propto L ^{0.2}_i$ (see Section \ref{sec:method}).
Both the foreground and the background (hidden) galaxy, GX, close to image C are also taken into account.
These two line-of-sight mass structures are modeled at the cluster redshift, as a multi-plane lensing framework is not yet fully implemented in the \texttt{lenstool} pipeline \citep[][]{Jullo2007}.
The two objects are then parametrized with a dPIE density profile, with their values of velocity dispersion and cut radius individually optimized in the lens model (i.e. outside of the scaling relations followed by the cluster members).

We summarize hereafter the differences between the two models.\\

\textit{- Model 1:}\\
This model includes an external shear component, which brings the total number of free parameters related to the mass parametrization to 21. 
The external shear component (corresponding to a non-localized constant) helps to significantly improve the reproduction of the multiple images (see Section \ref{sec:results} and Table \ref{tab:BFmod1}). The origin and impact of the external shear component is further discussed in Section \ref{sec:ExtShear}.\\

\textit{- Model 2:}\\
In this model, the two BCGs are individually parametrized (i.e. outside of the scaling relations): we consider them as spherical dPIEs with a vanishing core radius and only their velocity dispersion and cut radius values are optimized, within large flat priors.
This model has a total number of 23 free parameters related to the mass parametrization.\\

\section{Results and Discussion} \label{sec:results}
In this Section, we present the results of the two model optimizations. We also discuss the implications of adopting two different cluster total mass parametrizations on the model-predicted flux ratios and time delays between the quasar multiple images.

We show in Figure \ref{fig:dxdy} the separations along the $x$ and $y$ axes between the model-predicted and observed image positions. We find that both models reproduce accurately the observed positions of the multiple images with a resulting best-fit rms value of $0\arcsec.15$($0\arcsec.22$) for Model~1(2).

\subsection{Mass distribution} \label{sec:TMD}
As described above, the cluster total mass distribution is composed of two cluster-scale dark-matter halos centered around the two BCGs. 
The resulting median values of the parameters of the SL mass models, and the associated $1\sigma$ errors, are summarized in Table \ref{tab:BFmod1}. The values of the statistical estimators introduced in Section~\ref{sec:method} are also quoted.
The values of the parameters of the two SL models are overall consistent within the uncertainties, and considering the values of all figures of merit, Model 1 is favored. 
In addition, the models predict for system 1 a redshift value of $z_{\rm S1}=2.17_{-0.03}^{+0.02}$ for Model 1 and $z_{\rm S1}=2.14_{-0.03}^{+0.02}$ for Model 2. These values are in agreement with each other, and with the tentative QF~=~1 redshift measurement provided by MUSE (see Table \ref{tab:multim}). 
We have checked that the difference in the resulting values of the positions, velocity dispersions and core radii of the two cluster-scale halos in the two models can be explained by the degeneracy between the mass of the cluster-scale halos and of the two individually optimized BCGs in Model 2. In detail, we have found that for the two different models the cumulative mass values projected within circular apertures of 100~kpc ($\sim15\arcsec$) centred on the two BCGs are consistent, within the statistical uncertainties, despite the quite different contributions of the two most important mass components (i.e. the BCGs and the large-scale DM halos).

\begin{figure}
\centering
\includegraphics[width=\columnwidth]{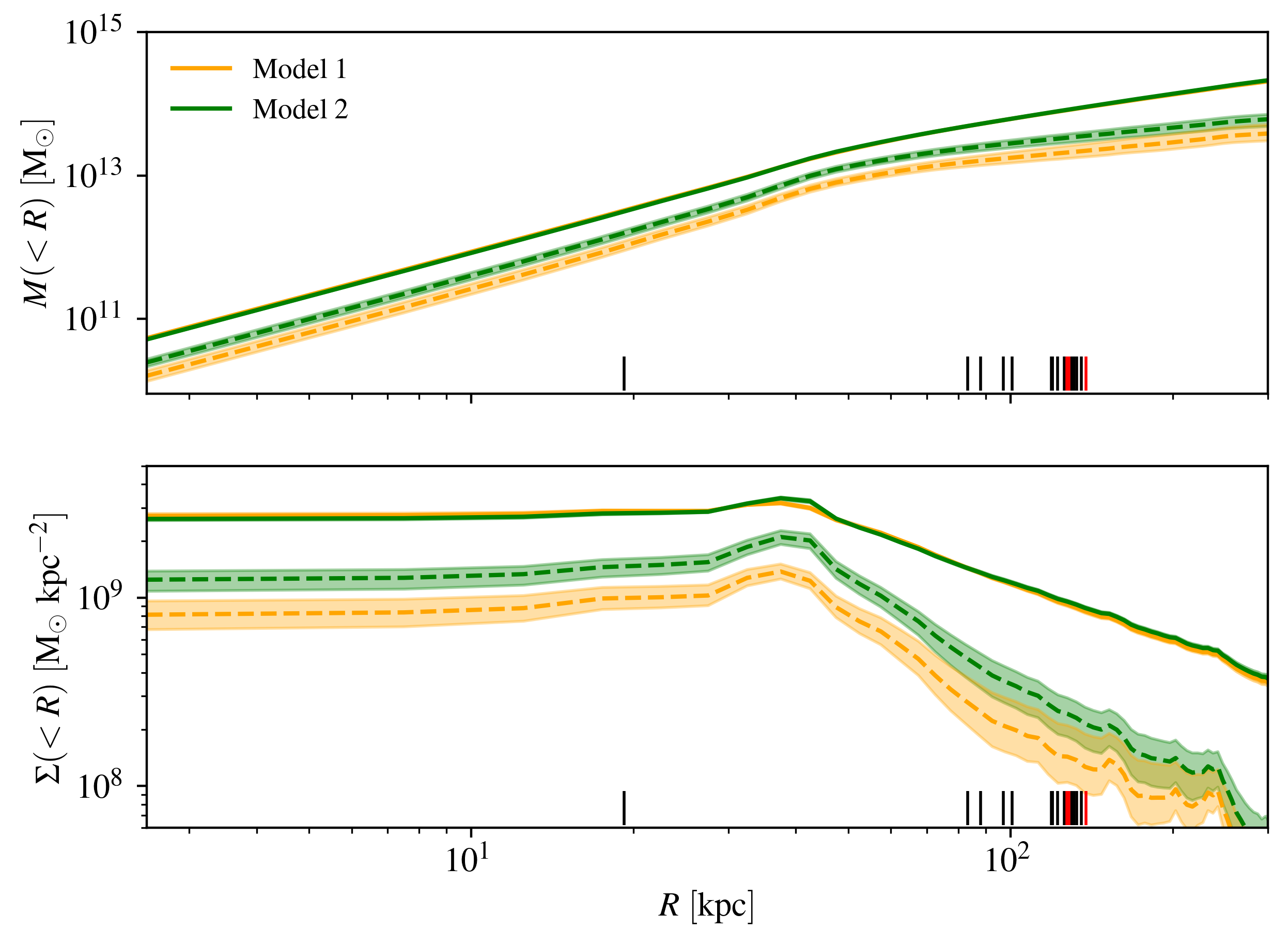} 
\caption{\textbf{Top:} Cumulative projected mass profiles.  \textbf{Bottom}: Average surface mass density profiles.
All profiles are plotted as a function of the distance ($R$) from the center of the cluster. The center is fixed at R.A.=157.303770\textdegree, Decl.=26.392601\textdegree (marked as a cyan cross in Figure \ref{fig:multim}) for the two SL models presented in this work.
The colored solid lines show the median values of the total mass and surface mass density profiles, while the dashed lines correspond to the mass component associated to the sub-halos.
The shaded areas encompass the 16th and the 84th percentiles, estimated from 500 random Bayesian Markov Chain Monte Carlo (MCMC) realizations.
The projected distances of the 26 multiple images from the cluster center are marked with vertical black lines and the distances of the three images of the quasar are highlighted in red.
} \label{fig:massprof}
\end{figure}

We show in Figure \ref{fig:massprof} the cumulative projected total (top panel) and the average surface mass density (bottom panel) profiles (not taking into account the two line-of-sight galaxies) as a function of the distance from the center of the cluster for both Model 1 (orange) and Model 2 (green). We also show the contribution from the sub-halo mass component as dashed lines.
We measure a precise projected total cluster mass of $M(<300~\rm kpc)= 2.04_{-0.03}^{+0.04} \times 10^{14}~ M_{\odot}$ for Model 1 and $M(<300~ \rm kpc)= 2.11_{-0.03}^{+0.04} \times 10^{14}~ M_{\odot}$ for Model 2. We find that these mass estimates are in agreement within the errors, and that the statistical uncertainty is approximately $2\%$. 
While the total cluster mass is precisely measured, the mass profile associated with the sub-halo component is significantly different between the two models. 
At a projected distance of 300 kpc from the cluster center, the sub-halo mass component represents approximately 20\% and 30\% of the cluster total mass for Models 1 and 2, respectively.
The difference is driven by the different modeling of the two central BCGs, and therefore the resulting values of their velocity dispersions. Indeed, the mass associated with the cluster members modeled within the scaling relations is consistent between the two models within the $1\sigma$ uncertainties (as shown in Figure \ref{fig:VELDIS}). The average surface mass density profiles shown in the bottom panel appear flat as they are computed from a reference center chosen between the two large-scale DM clumps (see Figure \ref{fig:multim}). We observe the same trend as for the mass profile, where the density profile associated with the sub-halo component is significantly different between the two models.

As displayed in Figure \ref{fig:masscon}, we find an overall agreement between the two models on the derived total surface mass-density distribution of SDSS1029. The differences can be explained by the different modeling of the BCGs in the most central regions and the inclusion of an external shear field in Model 1, which is discussed below.

\subsection{On the origin and impact of the external shear field} \label{sec:ExtShear}
As mentioned in Section \ref{sec:MassMods}, Model 1 requires a non-negligible external shear component (i.e., $\gamma_{\rm ext}=0.09$) to reproduce well the multiple image positions.
Including an external shear term in SL models has proved to significantly improve the goodness of the reconstruction in specific cases \citep[for a comparison of several different cluster mass parametrizations, see][]{Caminha2019}, e.g. in MACS J0329.7$-$0211 and RX J1347.5$-$1145 \citep[$\gamma_{\rm ext}=0.07$ and $\gamma_{\rm ext}=0.10$, respectively,][]{Caminha2019}, MACS J1206.2$-$0847 \citep[$\gamma_{\rm ext}=0.12$,][]{Bergamini2019, Caminha2016}, Abell 370 \citep[$\gamma_{\rm ext}=0.13$,][]{Lagattuta2019}, and Abell 2744 \citep[$\gamma_{\rm ext}=0.17$,][]{Mahler2018}. This additional component can account for some significant lensing effects that would otherwise not be represented in the SL models. There are several possible reasons to include an external shear term in the SL modeling of SDSS1029.

\begin{figure}
\centering
\includegraphics[width=\columnwidth]{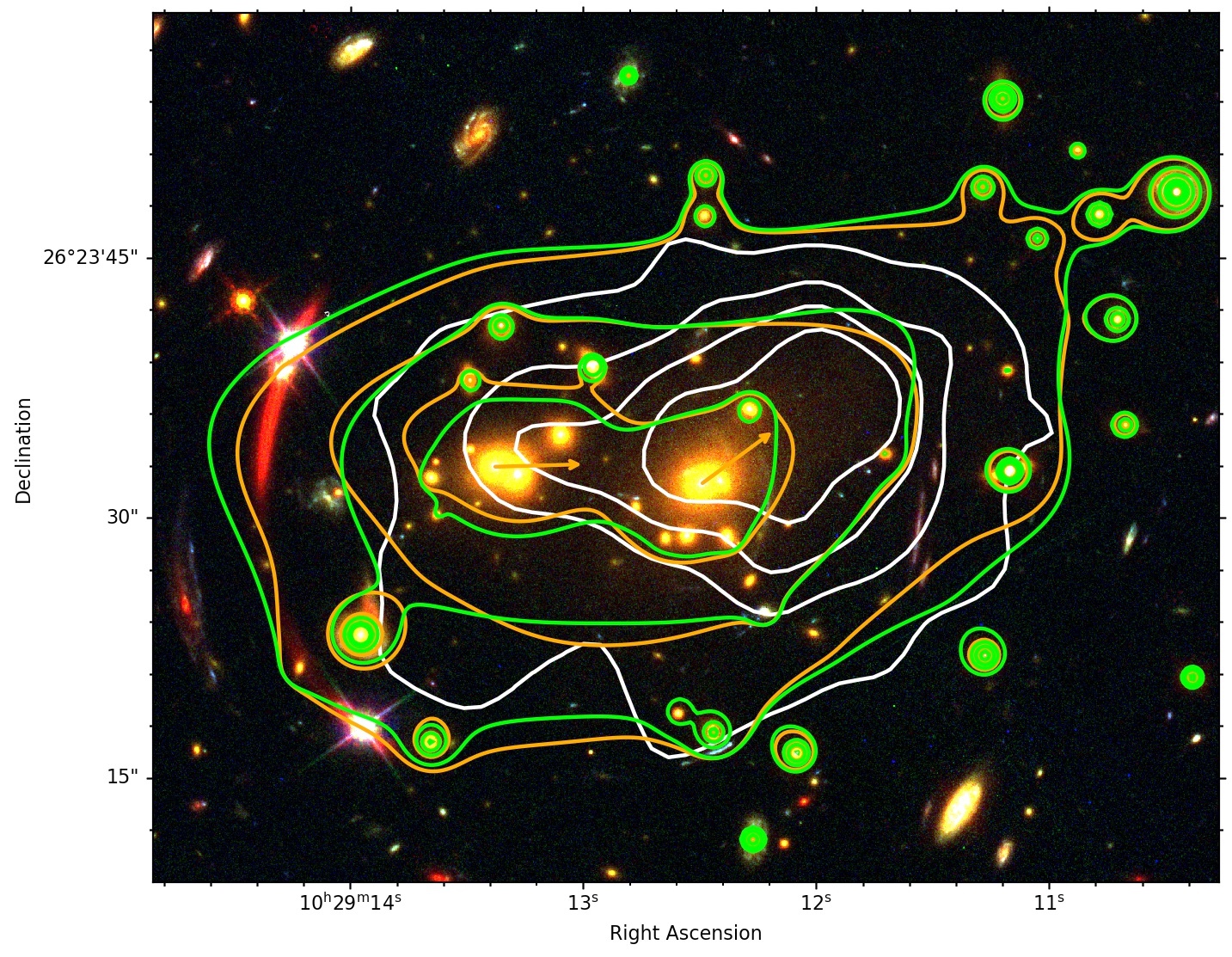}
\caption{Color composite image of the central region of SDSS1029, where we compare the contour levels of the total surface mass density distribution and the Chandra X-ray surface brightness (in white, after masking the strong emission from the three QSO images). Orange and green contours correspond, respectively, to the best-fit Models 1 and 2 and the levels are $[1.0, 1.5, 2.5] \mathrm{~ \times ~ 10^{9} ~ M_{\odot} ~ kpc^{-2}}$. The orange arrows indicate the orientation of the large-scale DM halos in Model 1.} \label{fig:masscon}
\end{figure}

SDSS1029 is a complex galaxy cluster with two main merging sub-clumps (see e.g. Figure~\ref{fig:masscon}). 
The X-ray surface brightness obtained from Chandra observations (Observation ID: 11755) is centered near one of the brightest cluster galaxies (ID 1933 in Table \ref{tab:BFmod1}) and is elongated in the East–West direction \citep{Ota2012}. This is illustrated in Figure \ref{fig:masscon}, where we show the X-ray surface brightness contours obtained from the smoothed total band image (0.5-7 keV) from which the signal from the quasar multiple images has been subtracted. The image is obtained with a minimal Gaussian smoothing of $\sigma=1\arcsec$. 
\citet{Ota2012} also identified a subpeak of X-ray emission northwest of the peak centered on the BCG with ID 1933, further supporting the merging scenario. Interestingly, we note that the large-scale DM halo associated with that BCG is oriented towards the X-ray emission subpeak in both SL models (see Figure \ref{fig:masscon}).
Recent studies have also found that massive structures at the redshift of a cluster but outside its central regions \citep[][]{Jauzac2016, Acebron2017, Mahler2018} or along the line-of-sight \citep[][]{McCully2017} can impact the mass reconstruction of the cluster core, an effect that can however vary significantly from cluster to cluster \citep{Chirivi2018}. This effect can be approximated with an external shear term.
Finally, the sub-halos associated with the cluster members are parametrized in both our models with circular mass density profiles. The external shear component could also account for localized perturbations from non-modeled elliptical mass distributions.

We remark that the difference between our two SL models (see Table \ref{tab:BFmod1}) in the derived ellipticity values of the two DM halos can be related to the external shear term. The inclusion of an external shear field in Model 1 results in rounder large-scale DM halos \citep[see also][]{Lagattuta2019}. Specifically, the DM halos associated with the BCG with ID 1933 have ellipticity values that are consistent, given the statistical uncertainties, while the DM halo associated with the BCG with ID 1953 has an ellipticity value $\sim20\%$ larger in Model 2 (i.e., when the external shear term is not present).

\subsection{On the multiply-imaged quasar system} \label{sec:QSO}
The three quasar images form the most interesting system lensed by SDSS1029.  
Both models reproduce well the positions of the quasar multiple images, with best fit rms values of $\sim0\arcsec.1$ for this system. We note that our current strong lensing models do only include the observed positions of the quasar multiple images as observational constraints. We delegate to future work the inclusion of the observed quasar image flux ratios and time delays and the detailed modeling of the surface brightness of the quasar host galaxy \citep{Suyu2010, Monna2015, Monna2017}.

\begin{figure*}
\centering
\includegraphics[width=\columnwidth]{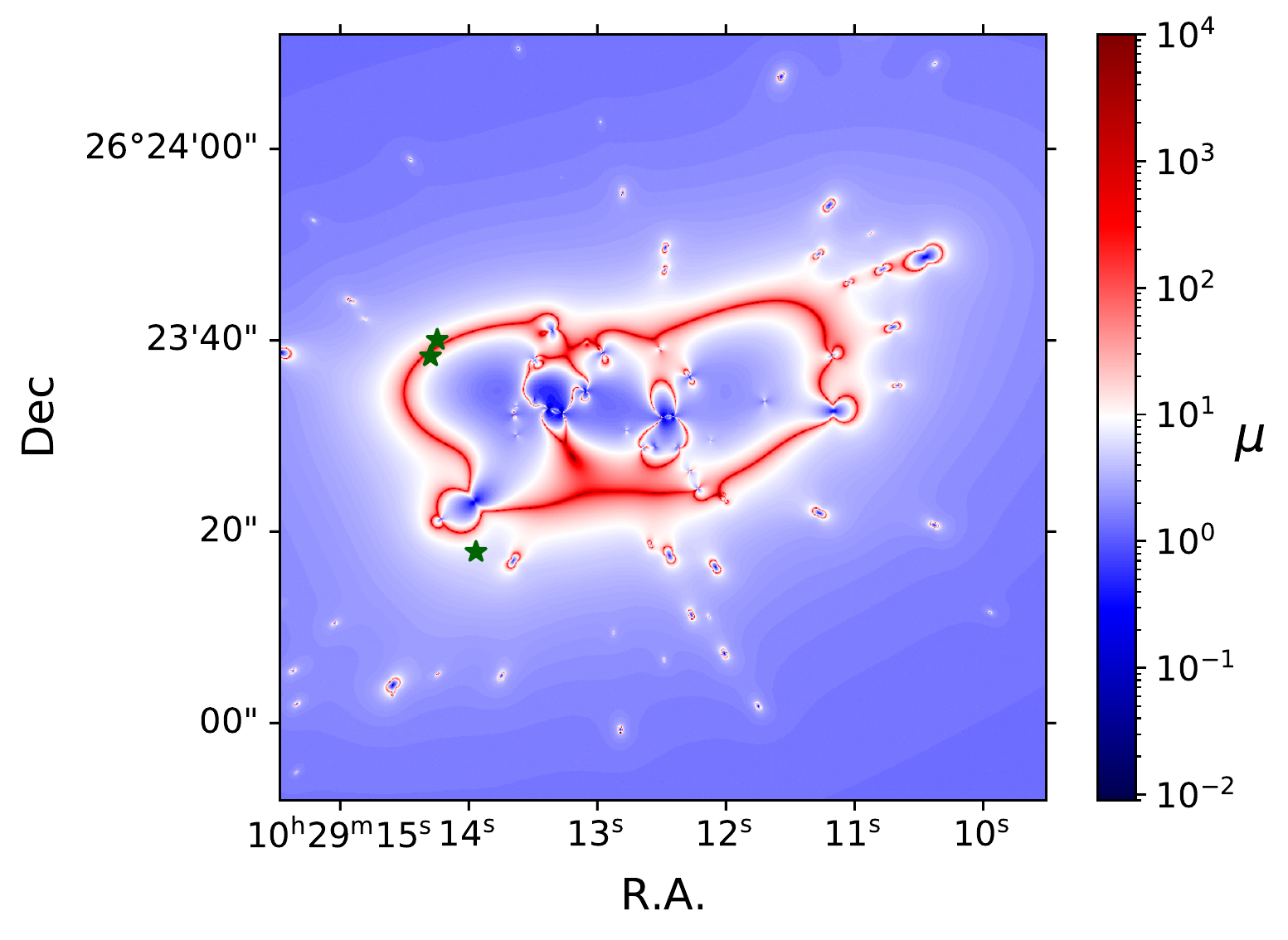}
\includegraphics[width=\columnwidth]{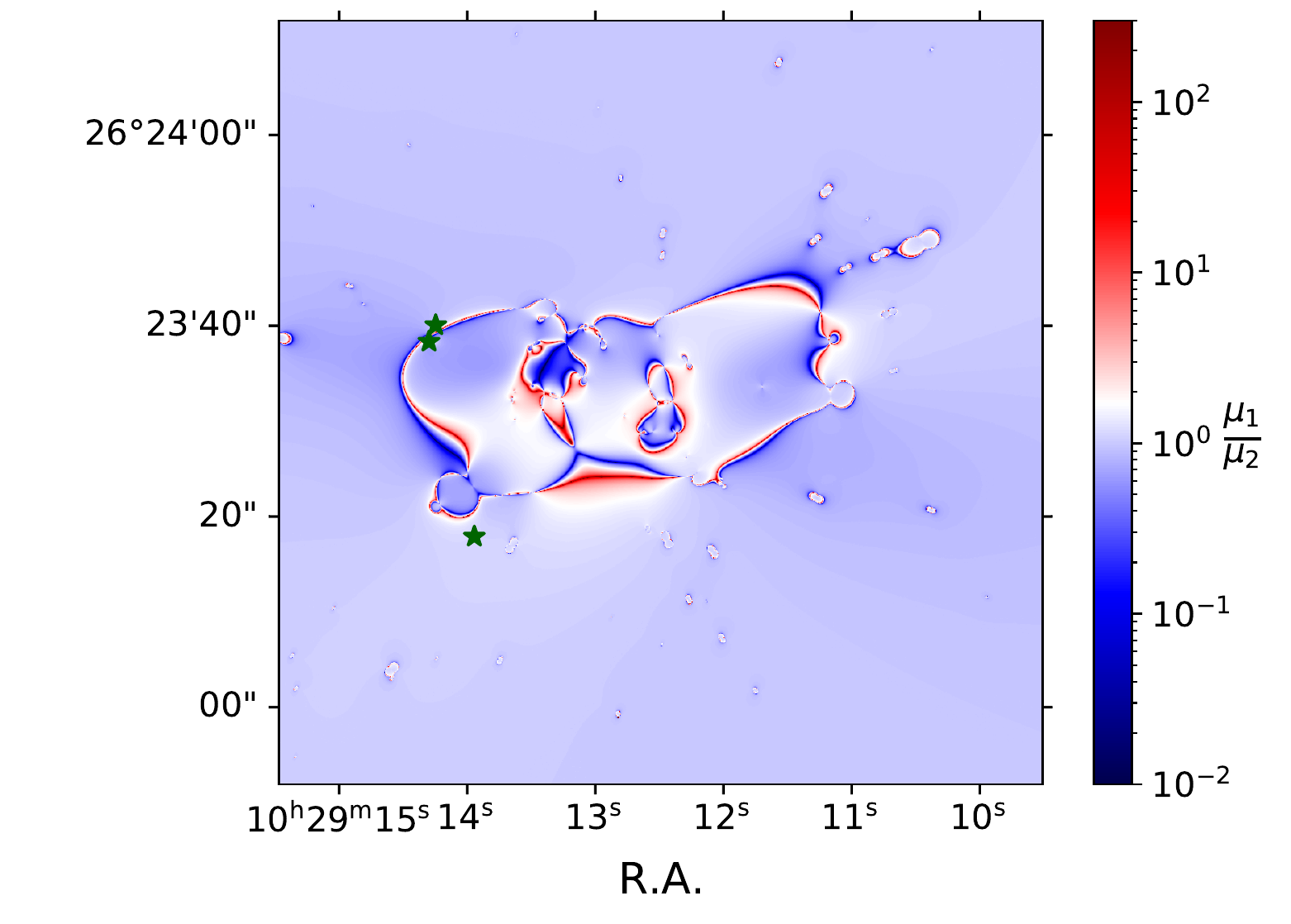} 
\caption{\textbf{Left:} Absolute magnification map of SDSS1029 from our best-fit Model 1 for a source at the quasar redshift $z = 2.1992$. \textbf{Right:} Magnification map ratio between the best-fit Model 1 and Model 2. In both panels, the green stars mark the observed positions of the three quasar images.} \label{fig:magzqso}
\end{figure*}
\begin{deluxetable*}{cccccc}
\tablenum{4}
\tablecaption{Values of model-predicted magnifications and model-predicted and observed magnitude differences between the three quasar images. \label{tab:fluxratios}}
\tablewidth{0pt}
\tablehead{
\colhead{} & \colhead{Model 1\tablenotemark{a}} & \colhead{Model 1 w/o GX\tablenotemark{a}} & \colhead{Model 2\tablenotemark{a}} & \colhead{Model 2 w/o GX\tablenotemark{a}} & \colhead{Observations\tablenotemark{b}} 
}
\startdata
$\mathrm{\mu_A}$ & $6.5_{-0.4}^{+0.4}$ & $6.7_{-0.5}^{+0.4}$ & $5.2_{-0.3}^{+0.4}$ &  $ 5.6_{-0.3}^{+0.3}$ &-- \\
$\mathrm{\mu_B}$ & $16.9_{-5.0}^{+6.4}$ & $23.8_{-2.6}^{+3.9}$ & $12.3_{-2.9}^{+5.8}$ & $24.6_{-3.5}^{+4.4}$ &-- \\
$\mathrm{\mu_C}$ & $5.6_{-3.0}^{+12.6}$ & $24.6_{-2.8}^{+4.1}$ & $3.0_{-1.2}^{+5.7}$ & $25.8_{-3.7}^{+4.5}$ &-- \\
\hline
\hline
$\mathrm{\Delta mag(AB)}$ & $1.0_{-0.3}^{+0.3}$ & $1.4_{-0.2}^{+0.2}$ & $0.9_{-0.2}^{+0.4}$ & $1.6_{-0.2}^{+0.2}$ & $0.36 \pm 0.02$ \\
$\mathrm{\Delta mag(BC)}$ & $-1.2_{-0.4}^{+1.1}$ & $0.0_{-0.1}^{+0.1}$ & $ -1.6_{-0.3}^{+0.9}$  & $0.0_{-0.1}^{+0.1}$ & $-1.96 \pm 0.05$  \\
$\mathrm{\Delta mag(AC)}$ & $-0.2_{-0.8}^{+1.2}$ & $1.4_{-0.2}^{+0.2}$ & $-0.6_{-0.5}^{+1.1}$ & $1.7_{-0.2}^{+0.2}$  & $-1.60 \pm 0.05$\\
\enddata
\tablenotetext{a}{The median magnification estimates and associated $1\sigma$ uncertainties are computed at the model-predicted positions of the multiple images from a subset of 500 MCMC random realizations.}
\tablenotetext{b}{Based on magnitudes in the $G$ band from the Gaia EDR3 catalog, which are mag(A)=18.67, mag(B)=18.31 and mag(C)=20.27.}
\end{deluxetable*}

This system is also of particular interest due to the observed ``flux ratio anomalies". As pointed out in \citet{Oguri2013}, preliminary lens models would predict similar magnitudes for the images B and C (which lie close together and close to the tangential critical curve, see Figure \ref{fig:multim}), while image A would appear fainter (i.e. $\mu_{\rm A}<\mu_{\rm B}\sim\mu_{\rm C}$). However, we observe $\mu_{\rm A}\sim\mu_{\rm B}>\mu_{\rm C}$.
We show in Table \ref{tab:fluxratios} the model-predicted magnification values (estimated at the model-predicted image positions) together with the observed and model-predicted magnitude differences between couples of quasar images for both models, computed as $\Delta {\rm mag}(XY) = -2.5\log\big(\frac{\mu_X}{\mu_Y}\big)$. 
The observed magnitudes are the $G$ band mean values from the Gaia EDR3 public catalog and we associate a typical error of 0.01 mag in the $G$ magnitude range $\approx$ 18-20 \citep[][]{Riello2021}, appropriate for the images A, B and C.
To quantify the impact of the background small galaxy GX, we also present the model-predicted magnification values and magnitude differences when the potential associated with this galaxy is not included in the models, therefore fully taking into account how a model would compensate for the absence of GX with the other cluster mass components.    
In Figure \ref{fig:magzqso}, we display the best-fit absolute magnification map computed at the quasar redshift, and the map ratio between the two models. 
The magnification of image A has a low relative error, as it is located far from the critical curves, and both models predict a magnification value lower than that of image B. 
On the other hand, the relative errors for images B and C are significantly larger since they are close to the tangential critical curve \citep{Meneghetti2017}. In particular, predicting a precise magnification estimate for image C is a complex undertaking, due to its closeness to both the critical curve and the background perturber.
Within the uncertainties, our magnification values are in agreement with the best-fit estimates quoted in \citet{Oguri2013} for their SL model including the hidden galaxy. 
We also find that the observed and model-predicted magnitude differences are consistent within $2\sigma$ for both models, with Model~2 being slightly in better agreement with the observations. 
In order to get a predicted magnitude difference between the images B and C ($\mathrm{\Delta mag (BC)}$) in agreement with the observations, the inclusion of the hidden background galaxy is crucial, comparable to the finding reported in \citet{Oguri2013}. In addition, \citet{Rojas2020} recently found no evidence of microlensing for this system. 
However, in contrast with the strong lensing model presented in \citet{Oguri2013}, we do not find that the inclusion of the GX substructure has an impact on the model-predicted magnification value for image A. GX has however a significant impact on the model-predicted magnitude differences between images A and B, and A and C (see Table \ref{tab:fluxratios}), resolving in part the observed ``anomaly". 
A possible scenario to explain the remaining differences between the model-predicted and observed values of $\mathrm{\Delta mag (AB)}$ and $\mathrm{\Delta mag (AC)}$ is the intrinsic variability of the source coupled with the time delays between the images. As shown by the light curves of images A and B+C combined (see Figure 4 in \citealt{Fohlmeister2013}), the background quasar appears as a significantly variable source, with a maximum $\Delta \rm mag \sim 0.7$ mag. Since the magnitudes of the multiple images of the quasar are measured from observations taken at the same time, but, because of the time delays, the corresponding emission times for the source are different, this could produce observed magnitude differences between the images of several tenths of magnitude.

Finally, we provide some discussion on the predicted time delays between the images A and B, $\Delta t_{\rm AB}$, and between the images B and C, $\Delta t_{\rm BC}$.
We use 500 random MCMC realizations from our SL models to estimate the median time delay between each couple of images (computed at the model-predicted image positions) and the $1\sigma$ associated errors.
We find time delay values of $\Delta t_{\rm AB}=740_{-210}^{+182}$ days and $\Delta t_{\rm BC}=5_{-3}^{+3}$ days for Model~1.
Regarding Model 2, we find $\Delta t_{\rm AB}=928_{-198}^{+200}$ days and $\Delta t_{\rm BC}=9_{-4}^{+6}$ days.
Both model predictions of $\Delta t_{\rm AB}$ are thus in agreement with the measured time delay of $744\pm 10$ days \citep{Fohlmeister2013} within the $1\sigma$ uncertainties.
We also find that not including the local perturber GX in the strong lensing models yields consistent, within the statistical uncertainties, $\Delta t_{\rm AB}$ values with respect to the fiducial Models 1 and 2. Only the model-predicted $\Delta t_{\rm BC}$ value significantly decreases with respect to the model-predicted values for Models 1 and 2, which can be explained by the fact that the light-rays traveling from the source to us are not pulled in the GX gravitational potential, and can reach us in a shorter amount of time.

It is worthy to note that two very similar cluster lens models (which estimate the same cluster total mass within $\sim2\%$, and reproduce equally well the positions of the quasar images) predict values of the $\Delta t_{\rm AB}$ time delay that differ by approximately $25\%$, which would in turn provide significantly different ${H_0}$ estimates. 
As previously mentioned, SDSS1029 presents a complex total mass distribution, with two merging mass components and a background perturber close to image C. As shown also in this work, perturbations introduced by substructures can have a significant impact on the predicted flux ratios and time delays between the quasar images \citep[see also][]{Oguri2007, Keeton2009}.
Furthermore, the quasar multiple images belong to the outermost strong lensing system identified in the field, in a region where no other multiple images are seen.

We remark that the results presented in \citet{Grillo2020} show that accurate and precise measurements of the values of the cosmological parameters through time delay cluster lensing require strong lensing models that include the measured values and uncertainties of the multiple image time delays as observational constraints. As proved also here, to be competitive with other cosmological probes, it is not possible to use or combine the time delay predictions obtained from different lens models that exploit only the observed positions of a set of multiple images (i.e. not including the measured time delays as observables).

Another interesting aspect in SDSS1029 comes from the quasar host galaxy which is lensed into a giant, $\sim27\arcsec$ long, tangential arc (see Figure \ref{fig:multim}). In a future work, we plan to go beyond \textit{point-like} SL models and reconstruct the surface brightness of the arc. This will grant us a much larger number of observational constraints, that will be used in combination with the measured time delays, to improve the accuracy and precision of our strong lensing models for proper cosmological applications.

\section{Conclusions} \label{sec:conclu}
In this work, we have presented a new parametric strong lensing model of the galaxy cluster SDSS J1029+2623 at $z=0.588$, which is one of the few currently known lens clusters hosting a multiply-imaged background quasar. We have exploited recent spectroscopic observations taken with MUSE in combination with archival HST imaging to securely identify multiple images and cluster galaxies.
We have used the new spectroscopic data to perform some accurate strong lens modeling of the cluster with the \texttt{lenstool} pipeline \citep{Jullo2007}.
Our main results are the following: 
\begin{enumerate}
    \item We have spectroscopically confirmed four multiple image systems, which were previously identified based solely on photometric information. We have provided an updated redshift for the quasar and a tentative redshift for one system, which is in excellent agreement with the redshift predictions from our strong lensing models. We have discovered a new system consisting of four multiple images at a redshift of $z=5.0622$, the highest redshift of our sample. In total, our SL models have included 7 families spanning a large redshift range between 1.0 and 5.1, with a total of 26 multiples images (see Table \ref{tab:multim} and Figures \ref{fig:multim} and \ref{specmi}).
    \item We have spectroscopically confirmed 63 cluster galaxies: 57 objects have been reliably identified with a QF $\geq$ 2 and 6 additional objects have a QF~=~1 and colors that make them reliable identifications. In addition, we have completed our spectroscopic catalog with a subset of member candidates outside the MUSE field of view, which were selected on the basis of their multi-wavelength photometric information (see Table \ref{tab:cm} and Figure \ref{figCM}).
    \item We have exploited the new MUSE spectra to measure the central stellar kinematics of a sub-sample of 20 cluster members down to F160W $\sim$ 21. The measured values of the stellar velocity dispersions have been in turn used to better constrain the scaling relations of the sub-halo population in the lensing models (see Figure \ref{fig:VELDIS}).
    \item We have confirmed the background nature of the hidden galaxy GX, angularly very close to image C of the quasar, at a redshift $z=0.6735$ (see Figure~\ref{fig:hiddengal}).
    \item We have presented two slightly different lens models that accurately reproduce the positions of all multiple images used as constraints, with rms values of the difference between the model-predicted and observed image positions of $0\arcsec.15(0\arcsec.22)$ for Model 1(2). In particular, the positions of the three quasar images are reproduced with a mean rms value of only $\mathrm{\sim0\arcsec.1}$ in both models.
    \item We have measured a cluster projected total mass value of $M\mathrm{(<300~kpc) \sim 2.1 \times 10^{14}~ M_{\odot}}$ for both models, with a relative statistical uncertainty of approximately $2\%$.
    \item We have confirmed, as previously stated in \citet{Oguri2013}, that the inclusion of the hidden galaxy GX in the lens model resolves the ``flux ratio anomaly" between images B and C. Those between images A and B/C instead could not be fully resolved with GX only. The intrinsic variability of the source, coupled with the  time delays between the images, might explain the remaining flux differences.
    \item Within the uncertainties, both strong lensing models predict time delays between images A and B that are consistent with the measured value presented in \citet{Fohlmeister2013}. We emphasize that these predictions should not be used to infer the value and uncertainty of the Hubble constant, since the measured time delay has not yet been included as an observational constraint in the models (which are optimized only with the observed positions of the multiple images).
\end{enumerate}

SDSS J1029+2623 is a complex merging cluster. In order to fully take advantage of the measured time delay between images A and B (with a remarkable $1\%$ relative error) for cosmological applications, all available lensing observables will be incorporated in future strong lens models. Towards this end, we plan to use the GLEE software \citep{Suyu2010,Suyu2012} that offers the opportunity to add as observational constraints in the SL modeling both the measured time delay $\Delta t_{\rm AB}$ and the surface brightness distribution of the multiple images of the quasar host galaxy.

The full MUSE spectroscopic catalog of SDSS~J1029+2623 presented here is made publicly available \footnote{The catalog is available at www.fe.infn.it/astro/lensing.}.

%% IMPORTANT! The old "\acknowledgment" command has be depreciated. It was not robust enough to handle our new dual anonymous review requirements and thus been replaced with the acknowledgment environment. If you try to compile with \acknowledgment you will get an error print to the screen and in the compiled pdf.

\begin{acknowledgements}
We kindly thank the referee for the insightful and useful comments received.
We acknowledge financial support through grant PRIN-MIUR 2017WSCC32 ``Zooming into dark matter and proto-galaxies with massive lensing clusters''.
A. A warmly thanks S. H. Suyu for her useful and constructive comments.
GBC thanks the Max Planck Society for support through the Max Planck Research Group for S. H. Suyu and the academic support from the German Centre for Cosmological Lensing.
\end{acknowledgements}


\vspace{5mm}
\facilities{HST(ACS,WFC3), VLT(MUSE)}


%\software{}

%\software{astropy \citep{2013A&A...558A..33A,2018AJ....156..123A},  
 %         Cloudy \citep{2013RMxAA..49..137F}, 
 %         Source Extractor \citep{1996A&AS..117..393B}
 %         }



\newpage
\appendix
\restartappendixnumbering

\section{Multiple images} \label{sec:A1}

\begin{figure*}
\centering
\includegraphics[width=\linewidth]{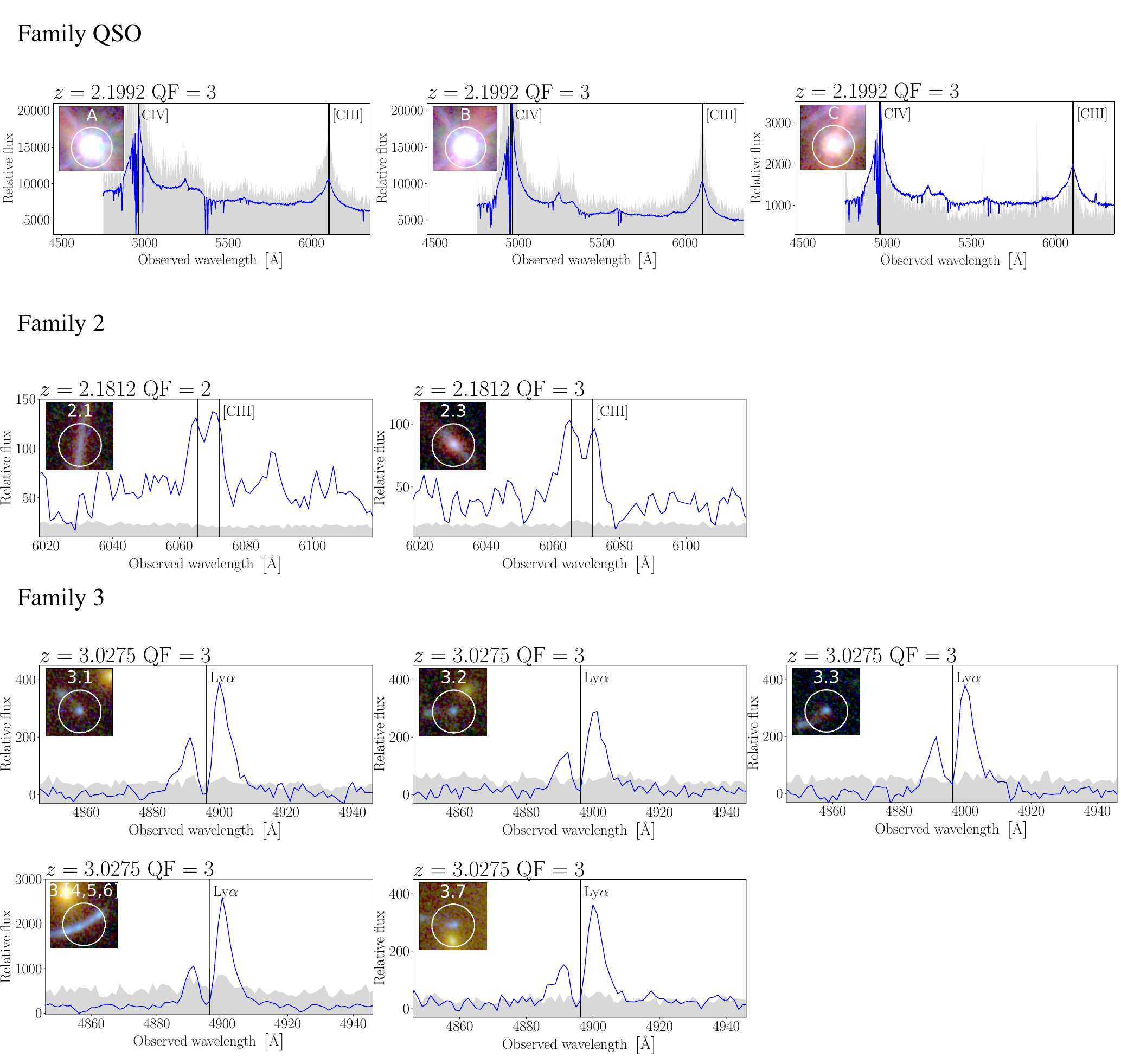} 
\caption{MUSE data of the multiply imaged background sources identified in SDSS1029. The vertical black lines indicate the positions of the emission lines based on the best estimate of the systemic redshift. The gray area shows the rescaled variance obtained from the data reduction pipeline. The flux is given in units of 10 $\rm erg~s^{-1}~cm^{-2}$ \r{A}$^{-1}$. The image cutouts in each panel are extracted from the color-composite HST image and are $2\arcsec$ across. The white circles show the HST counterparts.} \label{specmi}
\end{figure*}

\begin{figure*}
\setcounter{figure}{\value{figure}-1}
\centering
\includegraphics[width=\linewidth]{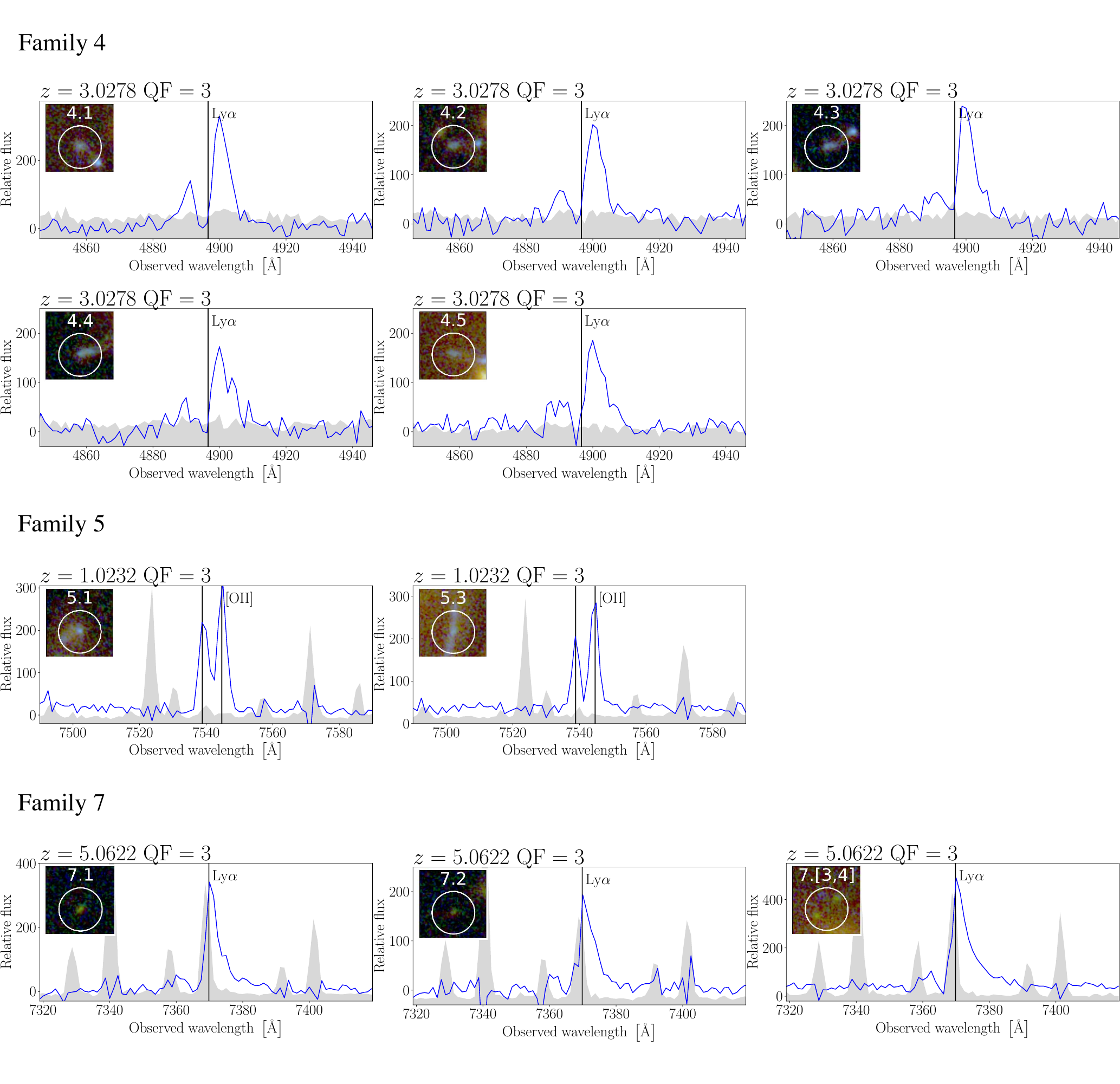} 
\caption{Continued}
\end{figure*}

\newpage

\section{Foreground and background sources from MUSE} \label{sec:A2}

\begin{longdeluxetable}{ccccc}
\tablenum{5}
\tablecaption{Catalog of the spectroscopically identified objects within the MUSE field-of-view pointed at SDSS1029. We present foreground and background objects in the top and bottom panels, respectively.
\label{tab:fgbgmuse}}
\tablewidth{0pt}
\tablehead{
\colhead{ID} & \colhead{R.A.} & \colhead{Decl} & \colhead{$z_{\rm spec}$} & \colhead{QF} \\
\colhead{} & \colhead{(deg)} & \colhead{(deg)}  & \colhead{} & \colhead{}
}
\startdata
1241 & 157.297428 & 26.387029 & 0.4085 & 3\\
1261 & 157.313315 & 26.385537 & 0.3396 & 3\\
1546 & 157.307545 & 26.386366 & 0.3419 & 3\\
1606 & 157.300511 & 26.386937 & 0.3407 & 3\\
1625 & 157.308160 & 26.389774 & 0.5111 & 3\\
1674 & 157.300040 & 26.387435 & 0.3611 & 3\\
1688 & 157.296005 & 26.387573 & 0.4080 & 3\\
2056 & 157.312978 & 26.391023 & 0.2936 & 3\\
2082 & 157.308720 & 26.391958 & 0.5120 & 3\\
2207 & 157.308551 & 26.392061 & 0.5120 & 3\\
99991 & 157.305466 & 26.397866 & 0.4869 & 9\\
999912 & 157.307886 & 26.386664 & 0.3419 & 2\\
\hline
1299 & 157.3092241 & 26.3838706 & 1.0299 & 3\\
1301 & 157.306078 & 26.384555 & 1.0229 & 3\\
1360 & 157.307012 & 26.384687 & 0.9078 & 3\\
1362 & 157.294096 & 26.384484 & 0.9075 & 3\\
1368 & 157.294083 & 26.384561 & 0.9075 & 3\\
1381 & 157.306315 & 26.384649 & 1.0233 & 3\\
1391 & 157.303578 & 26.384776 & 1.0240 & 3\\
1394 & 157.295228 & 26.384820 & 0.9272 & 3\\
1460 & 157.298442 & 26.385533 & 1.9781 & 2\\
1528 & 157.296660 & 26.386191 & 1.0965 & 3\\
1621 & 157.310328 & 26.387028 & 0.8840 & 9\\
1665 & 157.312845 & 26.387381 & 4.2281 & 3\\
1685 & 157.303819 & 26.387522 & 1.0881 & 9\\
1802 & 157.304357 & 26.388654 & 1.1130 & 9\\
1978 & 157.313342 & 26.390708 & 1.0036 & 3\\
2006 & 157.311295 & 26.390369 & 1.9094 & 2\\
2031 & 157.294407 & 26.391291 & 0.9190 & 3\\
2453 & 157.311805 & 26.394246 & 0.6640 & 3\\
2518 & 157.312155 & 26.395054 & 0.7227 & 3\\
2571 & 157.310895 & 26.395766 & 1.8942 & 3\\
2585 & 157.306033 & 26.397789 & 0.9390 & 3\\
2594 & 157.294471 & 26.395405 & 0.9183 & 3\\
2610 & 157.300366 & 26.395570 & 0.9177 & 3\\
2684 & 157.311124 & 26.396271 & 3.0296 & 3\\
2688 & 157.311085 & 26.396321 & 3.0296 & 3\\
2773 & 157.312375 & 26.397311 & 3.6651 & 2\\
2875 & 157.308164 & 26.399206 & 0.8168 & 3\\
2952 & 157.311901 & 26.399437 & 1.0892 & 3\\
3010 & 157.311906 & 26.399559 & 1.0894 & 3\\
3020 & 157.301205 & 26.399645 & 4.6553 & 3\\
3054 & 157.311282 & 26.399976 & 1.0887 & 3\\
3069 & 157.305358 & 26.400140 & 1.0579 & 3\\
3128 & 157.304154 & 26.401516 & 0.6736 & 2\\
99996 & 157.308018 & 26.390477 & 0.9170 & 3\\
99997 & 157.296211 & 26.386627 & 4.4242 & 3\\
999913 & 157.309500 & 26.393997 & 0.6735 & 2\\
\enddata
\tablecomments{ All listed coordinates are estimated with \texttt{SExtractor} based on the F814W band as detection image.}
\end{longdeluxetable}

\newpage

\section{Cluster members} \label{sec:A3}

\begin{longdeluxetable}{cccccc}
\tablenum{6}
\tablecaption{Catalog of the spectroscopic (top) and photometric (bottom) cluster members included in the SL modeling of SDSS1029. \label{tab:cm}}
\tablewidth{0pt}
\tablehead{
\colhead{ID} & \colhead{R.A.} & \colhead{Decl} & \colhead{F160W} &\colhead{$z_{\rm spec}$} & \colhead{QF} \\
\colhead{} & \colhead{(deg)} & \colhead{(deg)} & \colhead{(mag)}  & \colhead{}& \colhead{}
}
\startdata
1933\tablenotemark{g} & 157.302047 & 26.392209 & $18.24 \pm 0.10$ & 0.5853 & 3\\ 
1953\tablenotemark{g} & 157.305754 & 26.392473 & $18.47 \pm 0.10$ & 0.5959 & 3\\ 
1849\tablenotemark{g} & 157.305350 & 26.392348 & $18.55 \pm 0.10$ & 0.6017 & 3\\ 
2567\tablenotemark{g} & 157.293560 & 26.396878 & $19.36 \pm 0.10$ & 0.5877 & 3\\ 
2170\tablenotemark{s} & 157.296543 & 26.392404 & $19.60 \pm 0.03$ & 0.5914 & 3\\ 
1936\tablenotemark{s} & 157.304575 & 26.392983 & $19.97 \pm 0.04$ & 0.5976 & 3\\ 
2336\tablenotemark{s} & 157.304008 & 26.394089 & $20.02 \pm 0.04$ & 0.5858 & 3\\ 
1964\tablenotemark{g} & 157.301731 & 26.392245 & $20.07 \pm 0.10$ & 0.5868 & 2\\ 
1286\tablenotemark{g} & 157.310804 & 26.384452 & $20.10 \pm 0.10$ & 0.5987 & 3\\ 
2705\tablenotemark{s} & 157.296677 & 26.398377 & $20.46 \pm 0.05$ & 0.5868 & 3\\ 
2321\tablenotemark{s} & 157.301195 & 26.393396 & $20.49 \pm 0.05$ & 0.5864 & 3\\ 
2404\tablenotemark{s} & 157.305649 & 26.394742 & $20.49 \pm 0.05$ & 0.5802 & 3\\ 
1833\tablenotemark{s} & 157.296993 & 26.389444 & $20.65 \pm 0.05$ & 0.5865 & 3\\ 
1655\tablenotemark{s} & 157.300360 & 26.387879 & $20.69 \pm 0.06$ & 0.5821 & 3\\ 
1645\tablenotemark{s} & 157.301848 & 26.388213 & $20.69 \pm 0.06$ & 0.5841 & 3\\ 
2090\tablenotemark{s} & 157.302317 & 26.391371 & $20.73 \pm 0.06$ & 0.5881 & 3\\ 
2335\tablenotemark{s} & 157.306216 & 26.393884 & $20.86 \pm 0.06$ & 0.5921 & 3\\ 
1414\tablenotemark{s} & 157.300071 & 26.385367 & $20.88 \pm 0.06$ & 0.5799 & 3\\ 
1692\tablenotemark{s} & 157.306902 & 26.388058 & $20.91 \pm 0.06$ & 0.5899 & 3\\ 
2076\tablenotemark{s} & 157.301588 & 26.391352 & $20.99 \pm 0.06$ & 0.5862 & 3\\ 
1399\tablenotemark{g} & 157.301142 & 26.386493 & $21.07 \pm 0.10$ & 0.5878 & 3\\ 
1337\tablenotemark{g} & 157.307290 & 26.384720 & $21.08 \pm 0.10$ & 0.5929 & 3\\ 
2424\tablenotemark{s} & 157.294615 & 26.394830 & $21.11 \pm 0.07$ & 0.5850 & 3\\ 
1253\tablenotemark{s} & 157.298980 & 26.383842 & $21.13 \pm 0.07$ & 0.5771 & 3\\ 
2701\tablenotemark{s} & 157.301987 & 26.397147 & $21.22 \pm 0.07$ & 0.6002 & 3\\ 
2141\tablenotemark{s} & 157.306888 & 26.392297 & $21.34 \pm 0.08$ & 0.5832 & 3\\ 
2680\tablenotemark{s} & 157.294950 & 26.396518 & $21.42 \pm 0.08$ & 0.5843 & 3\\ 
2788\tablenotemark{g} & 157.303364 & 26.398741 & $21.46 \pm 0.10$ & 0.5857 & 3\\ 
2685\tablenotemark{s} & 157.297031 & 26.396958 & $21.55 \pm 0.08$ & 0.5911 & 3\\ 
2561\tablenotemark{s} & 157.312173 & 26.395616 & $21.70 \pm 0.09$ & 0.5927 & 3\\ 
1945\tablenotemark{g} & 157.300932 & 26.390149 & $21.72 \pm 0.30$ & 0.5830 & 3\\ 
2315\tablenotemark{s} & 157.294471 & 26.393140 & $21.82 \pm 0.09$ & 0.5844 & 3\\ 
2653\tablenotemark{s} & 157.302008 & 26.396494 & $21.83 \pm 0.09$ & 0.5810 & 3\\ 
1505\tablenotemark{s} & 157.312688 & 26.386234 & $22.03 \pm 0.10$ & 0.5926 & 3\\ 
2078\tablenotemark{s} & 157.302696 & 26.391330 & $22.06 \pm 0.10$ & 0.6026 & 3\\ 
2303\tablenotemark{s} & 157.298767 & 26.392680 & $22.07 \pm 0.10$ & 0.5835 & 3\\ 
2903\tablenotemark{s} & 157.310235 & 26.399697 & $22.09 \pm 0.11$ & 0.5991 & 3\\ 
1791\tablenotemark{s} & 157.302474 & 26.388525 & $22.18 \pm 0.11$ & 0.5826 & 3\\ 
2258\tablenotemark{g} & 157.306200 & 26.392746 & $22.19 \pm 0.20$ & 0.5969 & 1\\ 
1390\tablenotemark{s} & 157.309339 & 26.384769 & $22.24 \pm 0.11$ & 0.5866 & 3\\ 
2415\tablenotemark{s} & 157.296582 & 26.394014 & $22.29 \pm 0.12$ & 0.5839 & 3\\ 
1847\tablenotemark{s} & 157.309245 & 26.389256 & $22.34 \pm 0.12$ & 0.5936 & 3\\ 
2658\tablenotemark{s} & 157.296050 & 26.396134 & $22.34 \pm 0.12$ & 0.5888 & 3\\ 
2032\tablenotemark{s} & 157.301190 & 26.390650 & $22.35 \pm 0.12$ & 0.5841 & 3\\ 
2421\tablenotemark{s} & 157.302160 & 26.394210 & $22.54 \pm 0.13$ & 0.5883 & 3\\ 
1417\tablenotemark{s} & 157.302030 & 26.385186 & $22.57 \pm 0.13$ & 0.5970 & 3\\ 
2828\tablenotemark{g} & 157.313370 & 26.397935 & $22.63 \pm 0.20$ & 0.5924 & 1\\ 
2181\tablenotemark{s} & 157.306794 & 26.391706 & $22.73 \pm 0.14$ & 0.5823 & 3\\ 
1324\tablenotemark{s} & 157.310794 & 26.384203 & $22.78 \pm 0.15$ & 0.5981 & 1\\ 
2801\tablenotemark{s} & 157.295337 & 26.397552 & $22.87 \pm 0.15$ & 0.5848 & 2\\ 
1931\tablenotemark{s} & 157.300054 & 26.389811 & $22.93 \pm 0.16$ & 0.5775 & 3\\ 
1553\tablenotemark{s} & 157.300576 & 26.386445 & $22.94 \pm 0.16$ & 0.5884 & 2\\ 
2186\tablenotemark{g} & 157.303233 & 26.391831 & $22.96 \pm 0.20$ & 0.5851 & 3\\ 
3107\tablenotemark{s} & 157.304071 & 26.400806 & $23.15 \pm 0.17$ & 0.5896 & 3\\ 
1512\tablenotemark{g} & 157.303667 & 26.385967 & $23.16 \pm 0.20$ & 0.5908 & 2\\ 
2550\tablenotemark{g} & 157.311720 & 26.395081 & $23.28 \pm 0.40$ & 0.5803 & 1\\ 
2468\tablenotemark{g} & 157.304550 & 26.394394 & $23.45 \pm 0.40$ & 0.5919 & 1\\ 
2288\tablenotemark{s} & 157.306805 & 26.392550 & $23.54 \pm 0.21$ & 0.5834 & 1\\ 
2173\tablenotemark{s} & 157.300509 & 26.391560 & $23.94 \pm 0.25$ & 0.5863 & 2\\ 
2220\tablenotemark{g} & 157.299563 & 26.392007 & $25.89 \pm 0.50$ & 0.5821 & 3\\ 
2973\tablenotemark{s} & 157.305364 & 26.399178 & $26.00 \pm 0.64$ & 0.5859 & 3\\ 
2979\tablenotemark{g} & 157.305288 & 26.399204 & $26.06 \pm 0.40$ & 0.5859 & 3\\ 
1958\tablenotemark{g} & 157.300062 & 26.389995 & $26.22 \pm 0.50$ & 0.5819 & 3\\
\hline
663\tablenotemark{s} & 157.298034 & 26.374652 & $21.59 \pm 0.08$ & - & -\\ 
1185\tablenotemark{s} & 157.307819 & 26.379904 & $21.92 \pm 0.10$ & - & -\\ 
1301\tablenotemark{s} & 157.323507 & 26.381836 & $20.97 \pm 0.06$ & - & -\\ 
1446\tablenotemark{s} & 157.313932 & 26.381922 & $22.60 \pm 0.13$ & - & -\\ 
1523\tablenotemark{s} & 157.303442 & 26.383164 & $21.17 \pm 0.07$ & - & -\\ 
1527\tablenotemark{s} & 157.320831 & 26.383690 & $19.92 \pm 0.04$ & - & -\\ 
1645\tablenotemark{s} & 157.313911 & 26.383909 & $21.62 \pm 0.09$ & - & -\\ 
1701\tablenotemark{s} & 157.314040 & 26.384864 & $21.58 \pm 0.08$ & - & -\\ 
1878\tablenotemark{s} & 157.286797 & 26.385906 & $22.97 \pm 0.16$ & - & -\\ 
1958\tablenotemark{s} & 157.291461 & 26.386564 & $22.29 \pm 0.12$ & - & -\\ 
1974\tablenotemark{s} & 157.314545 & 26.386778 & $22.73 \pm 0.14$ & - & -\\ 
2257\tablenotemark{s} & 157.293281 & 26.389096 & $20.96 \pm 0.06$ & - & -\\ 
2374\tablenotemark{s} & 157.324898 & 26.390041 & $22.01 \pm 0.10$ & - & -\\ 
2930\tablenotemark{s} & 157.314447 & 26.394108 & $19.99 \pm 0.04$ & - & -\\ 
3281\tablenotemark{s} & 157.285541 & 26.396919 & $21.61 \pm 0.09$ & - & -\\ 
4015\tablenotemark{s} & 157.293265 & 26.402481 & $21.84 \pm 0.09$ & - & -\\ 
4033\tablenotemark{s} & 157.306745 & 26.402921 & $22.20 \pm 0.11$ & - & -\\ 
4040\tablenotemark{s} & 157.295230 & 26.403995 & $19.69 \pm 0.04$ & - & -\\ 
4396\tablenotemark{s} & 157.304565 & 26.405718 & $22.40 \pm 0.12$ & - & -\\ 
4435\tablenotemark{s} & 157.298666 & 26.406534 & $21.02 \pm 0.06$ & - & -\\ 
\enddata
\tablenotetext{g}{Coordinates and F160W magnitude (and associated error) are measured with \texttt{Galfit}.}
\tablenotetext{s}{Coordinates and F160W magnitude (and associated error) are measured with \texttt{SExtractor}.}
\end{longdeluxetable}


\bibliography{sample631}{}
\bibliographystyle{aasjournal}



\end{document}